\documentclass[showpacs, oneside, twocolumn, prd, amsmath, amssymb, nofootinbib, superscriptaddress]{revtex4-1}

\usepackage{cases}
\usepackage{amsmath}
\usepackage{amssymb}
\usepackage{amsfonts}
\usepackage{dcolumn}
\usepackage{bm}
\usepackage{bbm}
\usepackage{graphicx}
\usepackage{xcolor}
\usepackage{array}
\usepackage{caption}
\usepackage{subcaption}
\usepackage[colorlinks,linktocpage,linkcolor=cyan,citecolor=cyan]{hyperref}
\usepackage{wasysym}
\usepackage{cancel}

\definecolor{darkgreen}{RGB}{21,138,6}

\def\l{\left}
\def\r{\right}
\def\ddd{\mathrm{d}}
\def\be{\begin{equation}}
\def\ee{\end{equation}}
\def\ba{\begin{eqnarray}}
\def\ea{\end{eqnarray}}
\def\bl#1\el{\begin{align}#1\end{align}}

\begin{document}
\title{The backreaction effect of sound speed resonance in DBI inflation.}
\author{Bichu Li}
\email{libichu@mail.ustc.edu.cn}
\affiliation{Deep Space Exploration Laboratory/School of Physical Science,
University of Science and Technology of China, 96 Jinzhai Road, Hefei, Anhui 230026, P.R. China}
\affiliation{CAS Key Laboratory for Researches in Galaxies and Cosmology/Department of Astronomy, School of Astronomy and Space Science, University of Science and Technology of China, 96 Jinzhai Road, Hefei, Anhui 230026, P.R. China}

\author{Chao Chen}
\email{iascchao@ust.hk}
\affiliation{Jockey Club Institute for Advanced Study, The Hong Kong University of Science and Technology, Clear Water Bay, Kowloon, Hong Kong, P.R. China}

\author{Bo Wang}
\email{ymwangbo@ustc.edu.cn}
\affiliation{Deep Space Exploration Laboratory/School of Physical Science,
University of Science and Technology of China, 96 Jinzhai Road, Hefei, Anhui 230026, P.R. China}
\affiliation{CAS Key Laboratory for Researches in Galaxies and Cosmology/Department of Astronomy, School of Astronomy and Space Science, University of Science and Technology of China, 96 Jinzhai Road, Hefei, Anhui 230026, P.R. China}

\begin{abstract}

We examine the backreaction effect of the enhanced small-scale scalar perturbations from the sound speed resonance (SSR) mechanism for primordial black hole formation in Dirac-Born-Infeld (DBI) inflation 
Within the perturbative regime, the backreaction effect of perturbations on the background dynamics can be described by an effective action after integrating out the perturbation sector. Starting with the effective field theory of a specific DBI inflation model that incorporates SSR, we obtain the one-loop effective action by integrating out the scalar perturbations at the quadratic level. 
Using the effective Friedmann equations derived from this one-loop effective action, we solve the Hubble parameter with backreaction and the effective perturbation dynamics on this background as well. 
Our numerical findings reveal that, for a viable parameter space, the backreaction effect results in a relative correction to the Hubble parameter of approximately $10^{-7}$, whereas the relative correction to the slow-roll parameter can vary between $-0.3$ and $0.1$, before gradually converging to $ 10^{-7}$.
Furthermore, our results show that the backreaction effect on SSR sound speed causes a slight reduction in the resonant peak of the curvature power spectrum, and the subsequent PBH formation predicted by the SSR mechanism remains almost unchanged.

\end{abstract}

\maketitle

\section{Introduction}

The conventional perturbative approach to the classical Einstein's field equation relies on the balance of the perturbative orders between geometry and matter sectors, and implicitly assumes the negligible interference among different orders. 
However, the intrinsic nonlinearity of Einstein's gravity inevitably allows linear cosmological perturbations around a homogeneous and isotropic Friedmann-Lema\^{\i}tre-Robertson-Walker (FLRW) Universe to affect on the background dynamics, and this backreaction effect of linear perturbations arise as second and higher orders in the perturbation expansion. It is well known that gravitational waves (GWs) could affect the dynamics of the background on which they are propagating, and thus one can define the effective energy-momentum tensor of GWs by expanding Einstein's field equation up to the second order and taking a spatial average~\cite{Brill:1964zz, Isaacson:1968hbi, Isaacson:1968zza, Abramo:1997hu, Giovannini:2019oii}. In addition, considerable attention has been given to the backreaction effect of scalar perturbations in relation to whether small-scale inhomogeneities have an impact on the global FLRW evolution,
which may give a new solution to cosmological coincidence problem, i.e., the acceleration epoch begins roughly at the same moment that non-linear structures form~\cite{Adamek:2017mzb, Schander:2021pgt}. Besides the above-mentioned classical approach to the cosmological backreaction problem, semi-classical and quantum approaches are also investigated~\cite{Schander:2021pgt, Hu:2020luk}.

Intuitively, the backreaction effect is expected to be proportional to the amplitude of perturbations. In the context of primordial black hole (PBH) formation within the inflationary Universe, the small-scale curvature perturbations are required to be amplified around seven orders of magnitude larger than the large-scale perturbations constrained by cosmic microwave background (CMB) radiation experiments. 
The amplification mechanism requires a deviation from the standard slow-roll inflation, at least for a period with a few e-folds that accounts for PBH formation~\cite{Sasaki:2018dmp, Carr:2020xqk, Escriva:2022duf}. So far, there are plenty of inflationary models that successfully enhance the small-scale primordial curvature perturbations during inflation  and the resulting overdense regions can collapse into PBHs, such as a non-attractor phase during inflation~\cite{Garcia-Bellido:2017mdw, Germani:2017bcs, Motohashi:2017kbs, Pattison:2017mbe, Di:2017ndc, Byrnes:2018txb, Passaglia:2018ixg, Xu:2019bdp, Tasinato:2020vdk, Fu:2020lob, Inomata:2021tpx, Kawai:2021edk, ZhengRuiFeng:2021zoz, Balaji:2022rsy, Qiu:2022klm,  Fu:2022ssq, Kasai:2022vhq, Fu:2022ypp, Pi:2022zxs, Cai:2023uhc}, 
parametric resonance~\cite{Cai:2018tuh, Chen:2019zza, Cai:2019bmk, Ashoorioon:2019xqc, Zhou:2020kkf, Chen:2020uhe, Cai:2021wzd}, 
additional contributions from isocurvature perturbations~\cite{Linde:2012bt, Braglia:2020eai, Liu:2021rgq,  Meng:2022low, Chen:2023lou, Ferrante:2023bgz, Ge:2023rrq}
and potentially large non-Gaussianity~\cite{Ezquiaga:2019ftu, Atal:2019erb, Figueroa:2020jkf, Cai:2021zsp, Cai:2022erk, Matsubara:2022nbr, Gow:2022jfb, Pi:2022ysn},
A natural question arises: if such enhanced small-scale scalar perturbations have a certain impact on the background dynamics, and further alter the efficiency of amplification of perturbations expected in the original models?

For the ultra-slow-roll inflation, the backreaction effect was investigated in Ref.~\cite{Cheng:2021lif} under two concrete models using Hartree factorization, and authors found that the backreaction can boost the curvature perturbation on superhorizon scales.
The lattice simulation of the axion-U(1) inflation model in Ref.~\cite{Caravano:2022epk} shows that the non-Gaussianity of curvature perturbation is sensitive to the backreaction. 
Reference~\cite{Yu:2023ity} considers the backreaction of a spectator field on an inflationary background using Hartree approximation.
Moreover, several recent works on the one-loop corrections from curvature perturbations~\cite{Inomata:2022yte, Kristiano:2022maq, Riotto:2023hoz, Choudhury:2023rks, Choudhury:2023vuj, Firouzjahi:2023btw, Franciolini:2023lgy, Cheng:2023ikq, Fumagalli:2023hpa} and tensor perturbations~\cite{Chen:2022dah,Ota:2022xni} show the potential importance of the one-loop corrections in the context of PBH formation. Reference~\cite{Gong:2022tfu} studied the backreaction in the hybrid inflation by perturbatively expanding the action up to the one-loop order and integrating out the perturbation to obtain an effective one-loop action from the path integral formalism. 
In this work, we investigate the backreaction effect of the sound speed resonance (SSR) mechanism which enhances small-scale curvature perturbations through the parametric resonance triggered by a time-oscillating sound speed of inflaton \cite{Cai:2018tuh}. 
Reference~\cite{Chen:2020uhe} has shown that the SSR mechanism can be realized in the framework of Dirac-Born-Infeld (DBI) inflation. 
Following the method presented in Ref.~\cite{Gong:2022tfu}, we integrate out scalar perturbations at the quadratic order using the path integral formalism, and derive the modified Friedmann equations that measure the backreaction effect of perturbations on the background dynamics. We will show the backreaction corrections on the Hubble parameter, slow-roll parameter and sound speed, and further investigate backreaction on the enhancement of power spectrum.

This paper is organized as follows. In Sec.~\ref{Sec2}, we briefly review the SSR realized in DBI inflation.
In Sec.~\ref{sec3a}, we calculate the effective Friedmann equations from the one-loop effective action and provide numerical results for background parameters including the Hubble parameter, slow-roll parameter, and also the sound speed of curvature perturbations. In Sec.~\ref{sec3b}, we use the results from Sec.~\ref{sec3a} to calculate the curvature power spectrum with backreaction. Finally, we summarize the results in Sec.~\ref{Conclusion}.

\section{Realization of SSR in DBI inflation}\label{Sec2}

In this section, we will review the realization of SSR within DBI inflation~\cite{Chen:2020uhe}.
In the original SSR mechanism \cite{Cai:2018tuh}, the sound speed of inflaton is assumed to be time-oscillating during a few e-folds, 
\be \label{ssr:cs}
c_s^2(\tau) = 1 -2 \xi \l[ 1-\cos(2k_*\tau) \r] ~,~ \tau>\tau_s ~,
\ee
where $\tau$ is the conformal time, $\xi$ and $k_*$ are the small amplitude and frequency of the sound speed oscillation, respectively. $\xi < 1/4$ is required such that $c_s^2$ is positively defined, and the oscillation begins at $\tau_s$, where $k_*$ is chosen to be deep inside the Hubble horizon with $|k_*\tau_s|\gg1$. Consequently, the resonance of the curvature perturbations happens inside the Hubble horizon and leads to PBH formation during the later radiation-dominated epoch.  

The DBI model contains a non-canonical kinetic term, which can be used to realize a non-trivial sound speed. The DBI action is written as
\be \label{ssr:dbi}
S_{\text{DBI}} =\int \ddd^{4} x \sqrt{-g} \l[ \frac{1}{f(\phi)} \l(1-\sqrt{1 - 2 f(\phi) X} \r) - V(\phi) \r] ~,
\ee
where $X \equiv - \frac12 g^{\mu \nu} \nabla_{\mu} \phi \nabla_{\nu} \phi$, and $f(\phi)$ is the so-called warp factor. The DBI action \eqref{ssr:dbi} is inspired by string theory with the inflaton field regarded as the radial position of branes moving inside a warped throat \cite{Born:1934gh, Silverstein:2003hf, Alishahiha:2004eh}, and it has rich cosmological phenomenology \cite{Cai:2010rt,Cai:2010wt,Cai:2008if,Cai:2010rt}. The sound speed in the DBI model can be expressed as
\be\label{sound_speed}
c_s^2 = 1 - f(\phi) \dot{\phi}^2 ~,
\ee
where the dot denotes the derivative with respect to the cosmic time. 

To realize a time-oscillating sound speed in the form of Eq.~\eqref{ssr:cs}, Ref.~\cite{Chen:2020uhe} introduces an oscillating term into the well-studied anti-de Sitter warp factor $f(\phi) = \lambda / \phi^4$ with $\lambda$ a constant, 
\be \label{ssr:warp}
f(\bar{\phi})= {\lambda \over \bar{\phi}^4} \big[ 1 - \Theta\l(\bar{\phi} - \bar{\phi}_s \r) C(\bar{\phi}) \big]  ~,
\ee
where $\bar{\phi}(\tau)$ is the homogeneous background of inflaton field $\phi$, and $\bar{\phi}_s \equiv \bar{\phi}(\tau_s)$ is evaluated at the beginning of sound speed oscillation. The oscillation function $C(\phi)$ is defined as
\be \label{C_factor}
C(\bar{\phi})
\equiv
\cos\l\{2 k_{*} \tau_{s} \exp \l[ { H (1 - \epsilon) \sqrt{\lambda} \over \sqrt{2 \xi}} \l( {1\over\bar{\phi}} - {1\over\bar{\phi}_{s}} \r) \r] \r\} ~.
\ee
where $H$ is the Hubble parameter. $\Theta\l(\bar{\phi}-\bar{\phi}_s \r)$ is the Heaviside step function, which divides the inflationary epoch into the non-oscillating ($c_s^2 = 1 -2 \xi$ for $\tau \le \tau_s$) and oscillating (i.e., Eq.~\eqref{ssr:cs}) phases. 
The functional form \eqref{C_factor} is derived from the background evolution of inflaton $\bar{\phi}(\tau)$,
\be \label{de_Sitter_phi0}
\bar{\phi}(\tau) \simeq \l( {1 \over \bar{\phi}_i} + {\sqrt{2 \xi} \over H(1-\epsilon) \sqrt{\lambda}} \ln \frac{\tau}{\tau_i} \r)^{-1} ~,
\ee
in a quasi de-Sitter spacetime $a(\tau) \simeq 1 /(\epsilon-1) H \tau$ where $\epsilon \equiv - \dot{H}/H^2$ is the slow-roll parameter. $\bar{\phi}_i$ is the value taken at the beginning moment of inflation $\tau_i$. Although this solution is solved by non-oscillating sound speed, it is considered applicable to both oscillating and non-oscillating phases \cite{Chen:2020uhe}.
The theoretical viability of this DBI model to realize SSR is also investigated in Ref.~\cite{Chen:2020uhe}.

\section{Backreaction in SSR-DBI inflation}
\label{Sec3}

\subsection{Backreaction effect on the background}\label{sec3a}

In order to obtain an effective background Lagrangian with backreaction, a partition functional consisting of a DBI inflaton and an external source $J(x)$ is introduced as
\be \label{action_1}
Z[J] = \int \mathcal{D} \phi  \exp\l[ i S_\text{DBI} + i \int \ddd ^{4}x \sqrt{-g(x)} J(x) \phi(x) \r] ~,
\ee
where $\int\mathcal{D}[\phi]$ is the functional integral over all possible inflaton field configurations. Using the semiclassical approach within the path integral formalism \cite{Burgess:2020tbq}, one can perturbatively expand $Z[J]$ in Eq.~\eqref{action_1} with Eq.~\eqref{ssr:dbi} around the classical background $ \bar{\phi}$, up to the quadratic order in terms of the field perturbation $\phi_{\rm per}$ (i.e., $\phi = \bar{\phi} + \phi_{\rm per}$. The first-order term of $Z[J]$ vanishes due to the classical equation of motion for $\phi_{\rm per}$).
Within the framework of the effective field theory (EFT) of inflation~\cite{Cheung:2007st}, the inflation field perturbation $\phi_{\rm per}$ is ``eaten" by graviton (namely the unitary gauge $\phi_{\rm per}=0$), one can write down the most general inflation action merely based on metric which respects the unbroken spatial diffeomorphism. After using the St\"{u}ckelberg trick, the time diffeomorphism can be restored and the quadratic action for the scalar degree of freedom of gravity is given by \cite{Cheung:2007st}
\be\label{EFT_action_2ndorder}
S^{(2)}
= \int \ddd t \ddd^3x a^{3}(t) 
\l[ \alpha (t) \dot{\pi}^{2} - \beta(t) { (\nabla\pi)^{2} \over a^{2}(t) } \r]  ~,
\ee
where the dot refers the derivative with respect to the cosmic time. $\pi$ corresponds to the broken time translation symmetry.  
The explicit forms of the coefficients $\alpha(t),\beta(t)$ in the above action can be obtained by matching the SSR-DBI model with EFT action (see Appendix \ref{app:EFT_SSR} for details). 
After integrating out the quadratic action \eqref{EFT_action_2ndorder} in the partition functional \eqref{action_1}, one can obtain the effective Lagrangian for background evolution as,
\be\label{totalLag}
\mathcal{L}_\text{tot} \equiv \mathcal{L}_\text{bg}  + \mathcal{L}_\text{br} ~,
\ee
which describes the backreaction effect and is calculated as follows (see Appendix~\ref{app:eff_action} for details),
\begin{widetext}
\bl \label{br:bglag}
\mathcal{L}_\text{bg} &= \frac12 M_{\mathrm{Pl}}^{2} R_\text{tot} + f(\bar{\phi}_\text{tot})^{-1} \l( 1 - \sqrt{1 - f(\bar{\phi}_\text{tot}) \dot{\bar{\phi}}_\text{tot}^2 } \r) - V(\bar{\phi}_\text{tot}) ~,
\\ \label{br:brlag}
\mathcal{L}_\text{br} &= {c_{s,\mathrm{bg}}^{-3}\over 128 \pi^2} \Bigg[ \Lambda_k^2 \l( \Lambda_k^2 - 2 \mathcal{M}^2 \r)
+ 2 \mathcal{M}^4 \ln\l(1 + {\Lambda_k^2 \over \mathcal{M}^2}\r) - 2 \Lambda_k^4 \ln \l( {\Lambda_k^2 + \mathcal{M}^2 \over \Lambda_M^2} \r) \Bigg]  ~,
\el
\end{widetext}
where we have used the subscript ``tot/bg" to represent the background with/without backreaction, and the subscript ``br" refers to the backreaction term arising from the integrated perturbation sector.
For example, the total background inflaton field is written as $\bar{\phi}_{\rm tot} = \bar{\phi}_{\rm bg} + \bar{\phi}_{\rm br}$, where $\bar{\phi}_{\rm bg}$ follows the evolution~\eqref{de_Sitter_phi0}.
In Eq.~\eqref{br:brlag}, $\Lambda_k \equiv c_{s,\mathrm{bg}}^{-1} H_\text{tot}$ is a physical cutoff at the sound horizon which is introduced to prevent the divergence of $Z[J]$ as Appendix \ref{app:eff_action} shows, and $H_\text{tot} \equiv H_\text{bg} + H_\text{br}$.
$\mathcal{M}^2 \equiv \frac14 \l( 3 - 2 s \r)^2 H_\text{tot}^2 - H_\text{tot} \dot s$ acts as an effective mass implicated by Eq.~\eqref{app:S2}, and $s \equiv \dot{c}_{s,\mathrm{bg}}/(H_\text{tot} c_{s,\mathrm{bg}})$ is a dimensionless parameter representing the rate of change of the sound speed during inflation in presence of the backreaction. 
$\Lambda_M^2$ is a parameter with the unit of the mass square \cite{Gong:2022tfu}, whose value is taken such that  $\mathcal{L}_\text{br}$ vanishes at the initial time of inflation $\tau_i$ when SSR and its backreaction have not occurred. Here we assume the backreaction correction on background is homogeneous \cite{Mukhanov:1996ak}, thus above parameters do not depend on the spatial coordinate.

From the total Lagrangian $\mathcal{L}_\text{tot}$ in Eq.~\eqref{totalLag}, we obtain the effective Friedman equations as follows,
\bl \label{hbg}
\dot{H}_\text{bg} = & - {1 \over 2 M_\text{Pl}^{2}} c_{s,\mathrm{bg}}^{-1}  \dot{\bar\phi}_\text{bg}^{2} ~, 
\\ \label{hbr}
\dot{H}_\text{br} = & {1 \over 2 M_\text{Pl}^{2}} \frac{\delta \mathcal{L}_\text{br} }{\delta \dot{\bar{\phi}}_\text{bg}} \dot{\bar{\phi}}_\text{bg} ~,
\el
where $\delta / \delta \dot{\bar{\phi}}_\text{bg}$ refers to the variation with respective to $\dot{\bar{\phi}}_\text{bg}$.

In principle, one can integrate the above equations to give the Hubble parameter with/without backreaction. However, since it involves the complicated integration of Eqs.~\eqref{C_factor} and \eqref{de_Sitter_phi0}, it is hard to investigate the effect of backreaction analytically. In the remaining part of this paper, we shall adopt the numerical method and analyze the modification of the SSR-DBI inflation background by the backreaction effect. 
As an illustration, we shall show the evolution of the Hubble parameter, slow-roll parameter, sound speed and power spectrum of curvature perturbations with and without backreaction for comparison.

For simplicity, we use $\bar{\phi}_{\rm bg}$ instead of $\bar{\phi}_{\rm tot}$ as the temporal parameter, using the relation $\dot{\bar{\phi}}_{\rm bg} = \sqrt{\frac{2 \xi}{\lambda}} \bar{\phi}_{\rm bg}^2$ indicated by Eq.~\eqref{de_Sitter_phi0}. According to Eqs.~\eqref{hbg} and~\eqref{hbr}, we obtain the approximation,
\be \label{num:hubb}
\frac{ \ddd  H_\text{tot}}{ \ddd  \bar{\phi}_\text{bg} } \simeq \frac{ \ddd  H_\text{bg}}{ \ddd  \bar{\phi}_\text{bg} } + \frac{1}{2 M_{\mathrm{Pl}}^2} {\delta \mathcal{L}_\text{br} \over \delta \dot{\bar{\phi} }_\text{bg} } ~.
\ee
On the right-hand side, the background term is given by \cite{Chen:2020uhe}
\begin{equation} \label{ssr:hdot}
{\ddd H_\text{bg} \over \ddd \bar{\phi}_{\rm bg} }
= -\sqrt{ {\xi \over 2 \lambda \l[ 1-2 \xi \l( 1-\Theta(\bar{\phi}_{\rm bg} -\bar{\phi}_{s}) C(\bar{\phi}_{\rm bg}) \r) \r] } } {\bar{\phi}_{\rm bg}^{2} \over M_\text{Pl}^{2}} ~,
\end{equation}
while the backreaction term is calculated as
\begin{widetext}
\be \label{br:varLag}
\begin{aligned}
\frac{1}{ 2 M_{\mathrm{Pl}}^2 } \frac{ \partial \mathcal{L}_\text{br} }{\partial \dot{ \bar{\phi} }_\mathrm{bg} }
& = \frac{c_{s,\text{bg}}^{-3}}{256 \pi^2 M_{\mathrm{Pl}}^2 \dot{\bar{\phi}}_\mathrm{bg}} \l[ - 4 \Lambda_k^2 + 4 \mathcal{M}^2 \ln\l(1 + \frac{\Lambda_k^2}{\mathcal{M}^2}\r) \r] \l( 8 s^2 H_\text{tot}^2 - 6 s H_\text{tot}^2 - 2 H_\text{tot} \dot s \r) c_{s,\text{bg}}^{-2} 
\\&
+  \frac{ 3 \l( c_{ s,\text{bg} }^{-2} - 1 \r) c_{s,\text{bg}}^{-3}}{256 \pi^2 M_{\mathrm{Pl}}^2 \dot{\bar{\phi}}_{\rm bg}} \l[ \Lambda_k^2 \l( \Lambda_k^2 - 2 \mathcal{M}^2 \r) + 2 \mathcal{M}^4 \ln\l(1 + \frac{\Lambda_k^2}{\mathcal{M}^2}\r) - 2 \Lambda_k^4 \ln \l( \frac{\Lambda_k^2 + \mathcal{M}^2}{\Lambda_M^2} \r) \r] ~,
\end{aligned} 
\ee
\end{widetext}
where $c_{s,\text{bg}}^2 = 1 - f(\bar{\phi}_{\rm bg}) \dot{\bar{\phi}}_\text{bg}^2$ based on Eq.~\eqref{sound_speed}.

\begin{figure}[ht]
    \centering
    \includegraphics[width=0.45\textwidth]{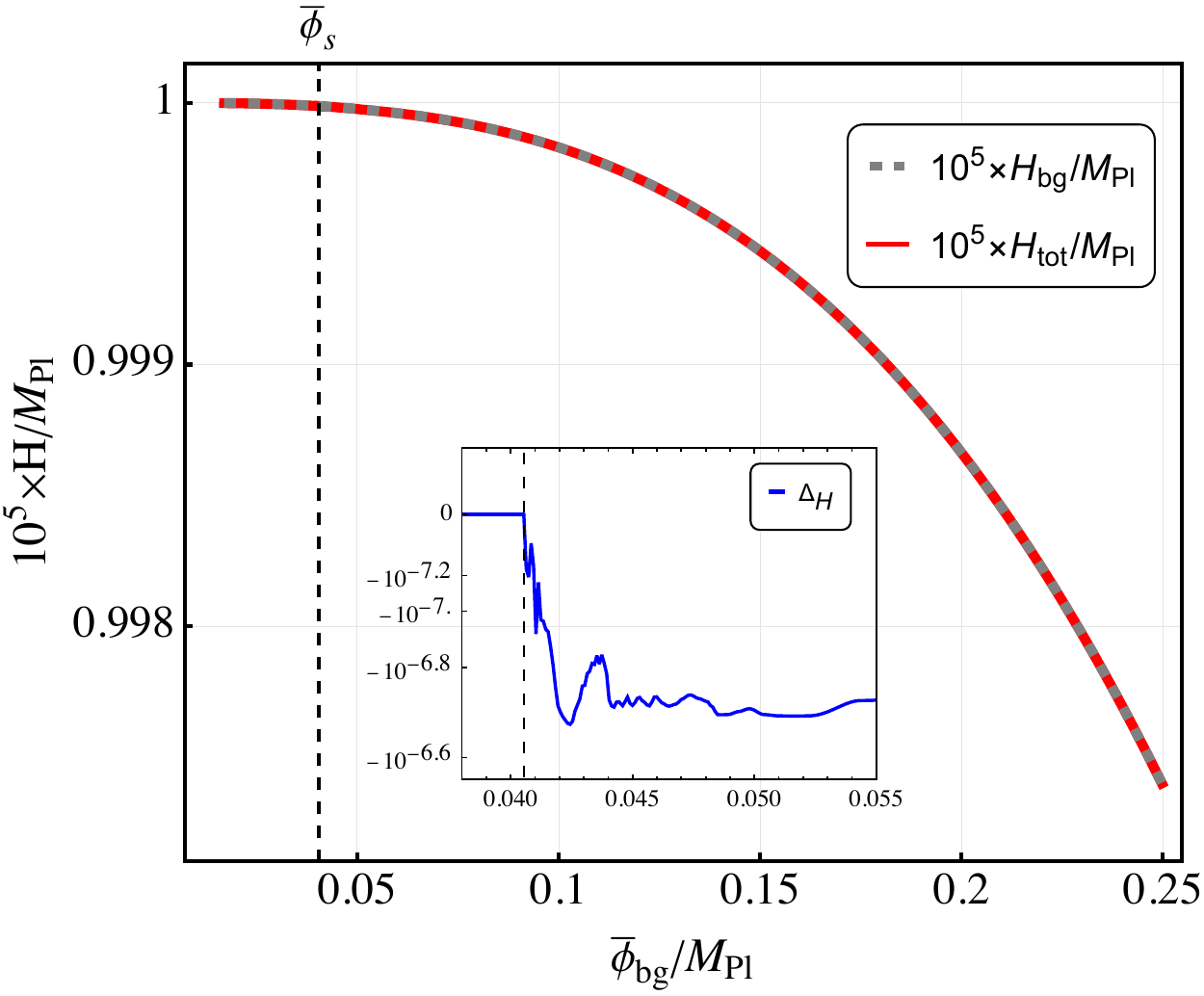}
    \caption{
    The evolutions of Hubble parameters $H_\text{tot}$ (red solid) and $H_\text{bg}$ (gray dashed) in terms of $\bar{\phi}_{\rm bg}$. 
    The black dashed vertical line refers to $\bar{\phi}_s$. The subplot shows the ratio of backreaction correction $\Delta_H \equiv H_\text{br}/H_\text{bg}$ around $\bar{\phi}_s$. The parameters are taken as~\cite{Chen:2020uhe}: $\lambda = 2 \times 10^9$, $\epsilon=10^{-3}$, $\xi=0.1$, $k_*=10$, $\tau_s=-14$, $N_s = 19.752$ and $H_i/M_{\mathrm{Pl}} = 10^{-5}$. 
}
    \label{fig:hubble}
\end{figure}
Numerically integrating Eq.~\eqref{num:hubb} gives the evolution of $H_\text{tot}$, which is shown by the red curve in Fig.~\ref{fig:hubble}. The ratio $\Delta_H \equiv H_\text{br}/H_\text{bg}$ is shown in the subplot by the blue curve.
The parameters taken in this paper follow Ref.~\cite{Chen:2020uhe}: $\lambda = 2 \times 10^9$, $\epsilon=10^{-3}$, $\xi=0.1$, $k_*=10$, $ \tau_s=-14$, $ N_s = 19.752$\footnote{ Here we choose a different $\tau_s$ with \cite{Chen:2020uhe} to obtain a curvature spectrum with a resonant peak $\mathcal{P}_\zeta (k_{peak}) \sim 0.01$ \cite{Green:2020jor}   } and $H_i/M_{\mathrm{Pl}} = 10^{-5}$ at the initial time $\tau_i$, then we calculate $\Lambda_M^2/M_{\mathrm{Pl}} \simeq 1.7 \times 10^{-5}$.
It is clearly seen from Fig.~\ref{fig:hubble} that $H_\text{tot}$ is fairly close to $H_{\rm bg}$ during the whole inflationary era, which demonstrates a negligible correction on the background expansion from the backreaction.
At the non-oscillating stage before SSR occurs ($\bar\phi_{\rm bg}<\bar\phi_{\rm s}$), the backreaction effect almost vanishes as expected, due to the tiny curvature perturbations. While during the oscillating stage, the backreaction grows to around $\Delta_H \sim 10^{-7}$ as a result of the enhanced curvature perturbations. 
Hence, we conclude that even if the peak of curvature perturbations are enhanced by around seven orders of magnitude through the SSR mechanism, their backreaction on the background expansion can be neglected. This is physically reasonable since only a narrow range of scalar mode functions around $k_*$ is enhanced in SSR, and their contribution to the background energy density comes from only a small fraction of the whole Fourier modes.

\begin{figure}[h]
    \centering
    \includegraphics[width=0.45\textwidth]{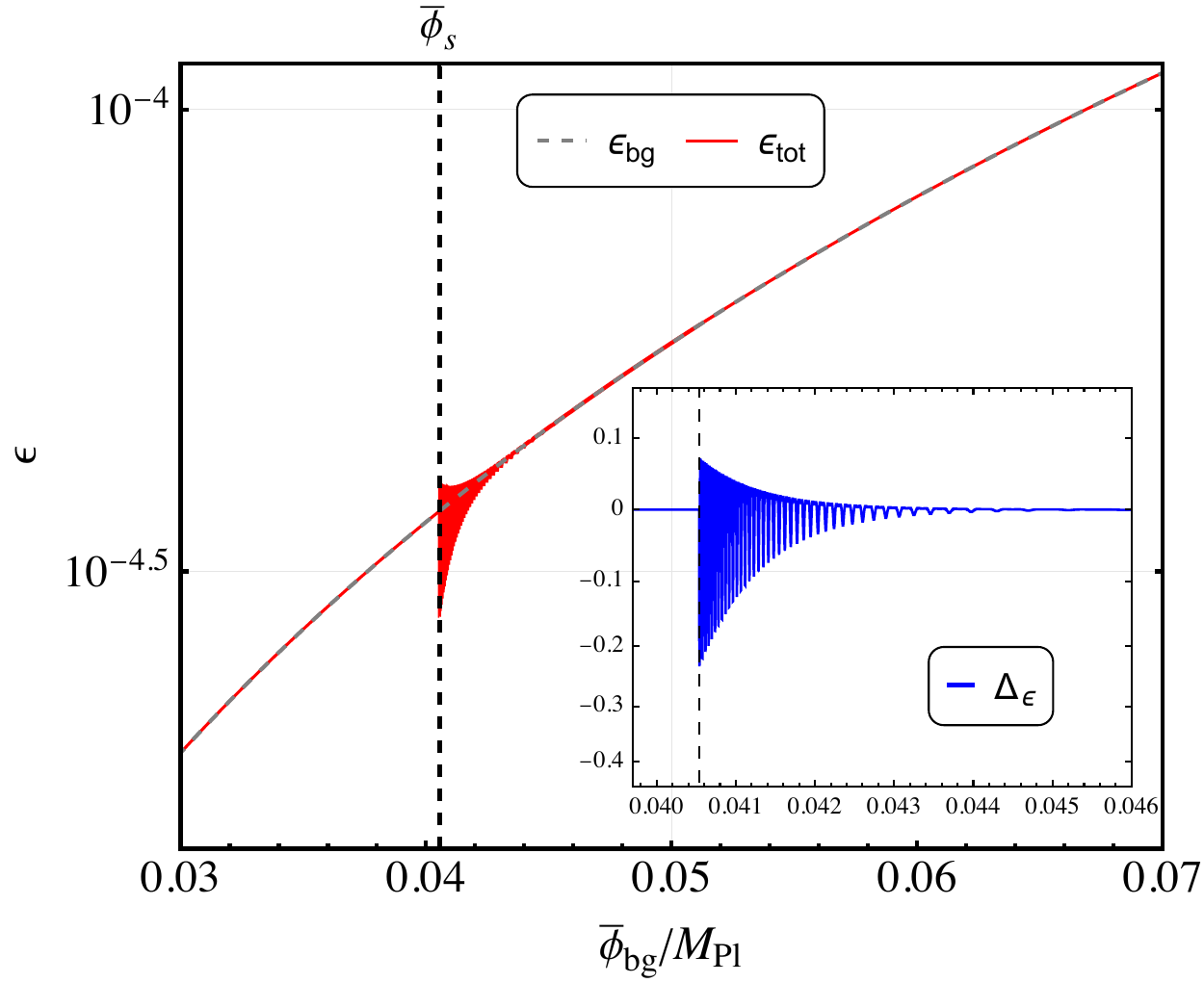}
    \caption{The evolutions of slow-roll parameters $\epsilon_\text{tot}$ (red solid) and $\epsilon_\text{bg}$ (gray dashed) in terms of $\bar{\phi}_{\rm bg}$. The ratio $\Delta_\epsilon \equiv \epsilon_\text{br}/\epsilon_\text{bg}$ is displayed in the subplot by the blue curve.}
    \label{fig:slowroll}
\end{figure}

\begin{figure}[h]
    \centering
    \includegraphics[width=0.45\textwidth]{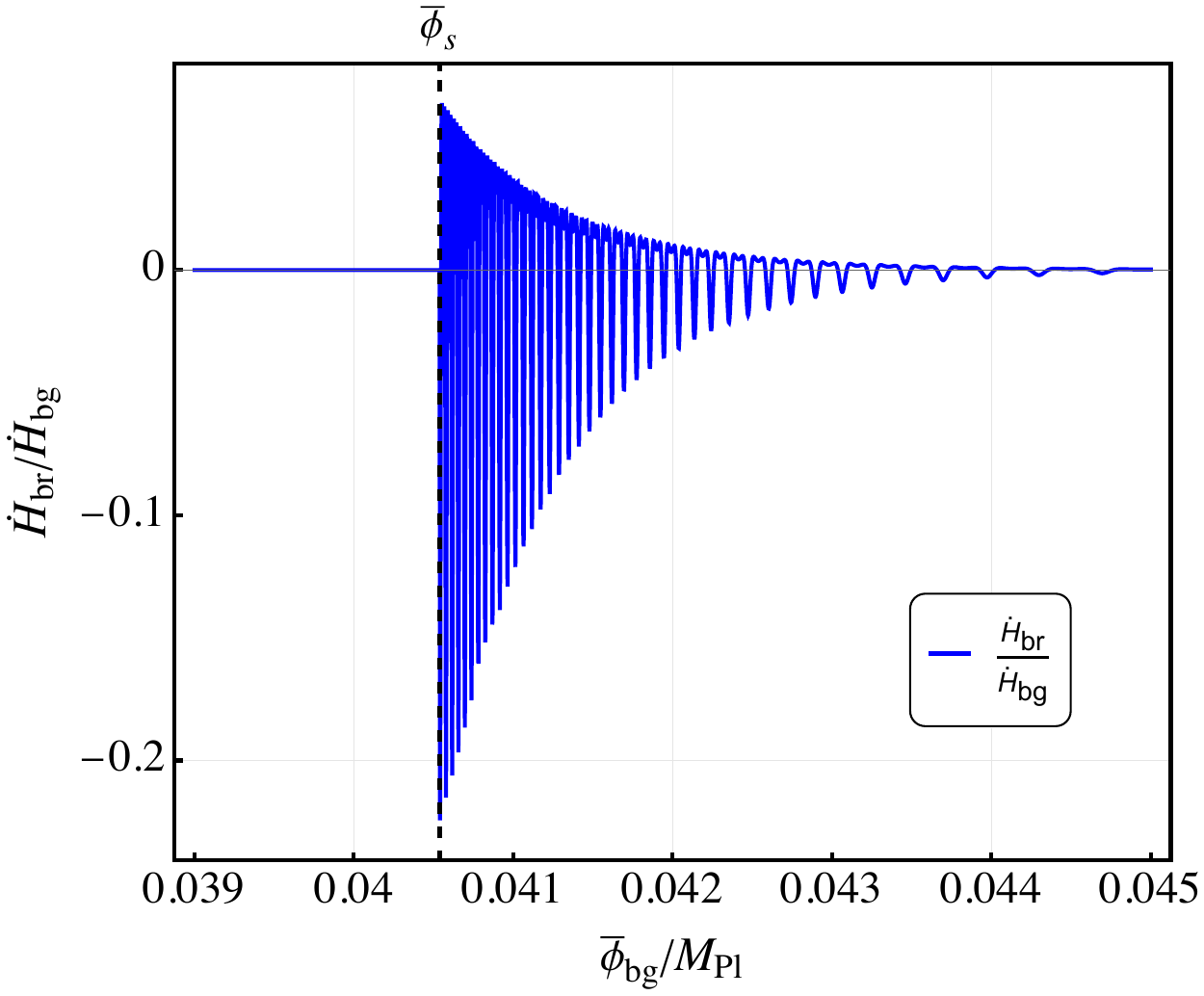}
    \caption{
    The evolution of $\dot{H}_\text{br} / \dot{H}_\text{bg}$ during SSR-DBI inflation. 
    }
    \label{fig:Hdot}
\end{figure}

Fig.~\ref{fig:slowroll}
shows the numerical result for the slow-roll parameter with/without backreaction defined as
$\epsilon_\text{tot} = - \dot{H}_\text{tot}/ H_\text{tot}^2$
and
$\epsilon_\text{bg} = - \dot{H}_\text{bg}/ H_\text{bg}^2$, and also shows the ratio of them defined as
\begin{equation} \label{def_delepsl}
\begin{aligned}
\Delta_\epsilon \equiv {\epsilon_\text{br} \over \epsilon_\text{bg} } 
& = 
{ 1 + \dot{H}_\text{br} / \dot{H}_\text{bg} \over (1 + \Delta_H)^2 } - 1 ~, 
\end{aligned}
\end{equation}
where $\epsilon_\text{br} \equiv \epsilon_\text{tot} - \epsilon_\text{bg}$.
The evolution of $\dot{H}_\text{br} / \dot{H}_\text{bg}$ is shown in Fig.~\ref{fig:Hdot}, which oscillates when $\bar \phi_{\rm bg}\gtrsim\bar\phi_s$, and then decays to  $\dot{H}_\text{br} / \dot{H}_\text{bg}\simeq - 10^{-7}\ll 1$ for larger $\bar{\phi}$, since $\dot{H}_\text{br} / \dot{H}_\text{bg} = \frac{\mathrm{d}}{\mathrm{d} t} \left( \Delta_H  H_\text{bg}  \right) / \dot{H}_\text{bg}$ and $\Delta_H \rightarrow -10^{-7}$ as shown in Fig.~\ref{fig:hubble}. 
Thus, we obtain
\begin{equation}\label{DeltaH}
\begin{aligned}
\Delta_\epsilon 
& \simeq - 2 \Delta_H = 2 \times  10^{-7}
~~~\text{ for large $\bar{\phi}_\mathrm{bg}$} ~,
\end{aligned}
\end{equation}
which is consistent with the numerical result in  Fig.~\ref{fig:slowroll}, where the backreaction effect on the slow-roll parameter experiences an oscillation at the beginning of the oscillating phase and then converges to zero.

Using the above results, we also investigate the backreaction effect on the background energy density $\rho = - T^0{}_{0} = - \left( \frac{\partial \mathcal{L}}{\partial\left(\partial_0 \phi_\alpha\right)} \partial_0 \phi_\alpha-g^0{}_0 \mathcal{L} \right)$
and pressure $P = \frac13 T^i{}_{i} = \frac13 \left( \frac{\partial \mathcal{L}}{\partial\left(\partial_i \phi_\alpha\right)} \partial_i \phi_\alpha-g^{i}{}_i \mathcal{L}\right)$.
By the decomposition of $\rho = \rho_\text{bg} + \rho_\text{br}$ and $P = P_\text{bg} + P_\text{br}$,
one writes the backreaction terms as follows,
\bl
\rho_\text{br} &= \rho_\text{bg} \l( 2 \Delta_H + \Delta_H^2 \r)  ~,
\\
P_\text{br} &= \rho_\text{bg} \Delta_H \l( {2\over3} \frac{\dot{H}_\text{br}}{ \dot{H}_\text{bg}} - \Delta_H - 2 \r) ~.
\el
Therefore, the equation of state of the backreaction term is  given by
\be\label{omega1}
\begin{aligned}
\omega_\text{br} \equiv {P_\text{br} \over \rho_\text{br}}
&= { \Delta_H \l( {2\over3} \dot{H}_\text{br}/ \dot{H}_\text{bg} - \Delta_H - 2 \r) \over 2 \Delta_H + \Delta_H^2 }  ~.
\end{aligned}
\ee
Since $\dot{H}_\text{br}/ \dot{H}_\text{bg} \ll 1 $ and $\Delta_H \ll 1$ as discussed above,
only considering leading terms,
Eq.~\eqref{omega1} becomes
\be\label{omega2}
\begin{aligned}
\omega_\text{br} \simeq -1 ~.
\end{aligned}
\ee
And since at leading order $\rho_\text{br} \simeq 2  \Delta_H \rho_\text{bg}$, where $\Delta_H \le 0$ according to the numerical result from Fig.~\ref{fig:hubble}, and $\rho_\text{bg} = \frac{3 H_\mathrm{bg}^2}{8 \pi G} >0$, thus one has $\rho_\text{br} < 0$. 
Therefore the backreaction effect of perturbation behaves like a negative cosmological constant, which is similar to the case of chaotic inflation~\cite{Brandenberger:2002sk}.

\subsection{Backreaction effect on the sound speed}

Next, we investigate the backreaction effect on the sound speed $c_s^2$ as the enhancement of the curvature perturbation $\zeta$ is sensitive to $c_s^2$ in SSR \cite{Cai:2018tuh, Chen:2020uhe}. 
In the next subsection, we shall further show how the power spectrum of $\zeta$ in the SSR-DBI inflation is influenced by the backreaction modification of $c_s^2$.
In this paper, we assume the functional form of the warp factor does not receive any correction from the backreaction, and the sound speed with backreaction in SSR-DBI inflation is thus given by
\be \label{total_cs}
c_{s,\text{tot}}^2 = 1 - f(\bar{\phi}_\text{tot}) \dot{\bar{\phi}}_\text{tot}^2 ~,
\ee
where $\bar{\phi}_\text{tot} (\tau)$ is written as
\be \label{phi_tot_dot}
\begin{aligned}
    \bar{\phi}_\text{tot} (\tau) \equiv & \left(\frac{1}{\bar{\phi}_i}+\frac{\sqrt{2 \xi}}{H_\text{tot}(1-\epsilon_\text{tot}) \sqrt{\lambda}} \ln \frac{\tau}{\tau_i}\right)^{-1} ~,
\end{aligned}
\ee
The derivative of $\bar{\phi}_\text{tot}$ under the slow-roll approximation $(\epsilon_\text{tot},~ \epsilon_\text{bg}) \ll 1$ and up to the first order of $\Delta_H$, is calculated as
\begin{equation}
\begin{aligned}
    \dot{\bar{\phi}}_\text{tot} (\tau) \simeq & 
    \left[ 1 + \left( 1 - 2 \frac{\bar{\phi}_\text{bg}}{\bar{\phi}_i} \right) \left( \Delta_H - \epsilon_\text{br} \right)  \right] \dot{\bar{\phi}}_\text{bg}  \\
    & + \left( 1 - \frac{\bar{\phi}_\text{bg}}{\bar{\phi}_i} \right) \bar{\phi}_\text{bg} \left( \dot{\Delta}_H - \dot{\epsilon}_\text{br} \right) ~.
\end{aligned}
\end{equation}
Notice that in Eq.~\eqref{num:hubb}, we do not consider the backreaction in $\bar{\phi}_{\rm tot}$ when calculating $H_{\rm br}$. In principle, one can use $\bar\phi_{\rm tot}$ to replace $\bar\phi_{\rm bg}$ to obtain a more precise prediction of $H_{\rm br}$.
More detailed research shall be carried out in our future work.

Based on the form of $\phi_\text{tot}$ in Eq.~\eqref{phi_tot_dot}, the oscillation term $C\left(\phi_\text{tot}(\tau)\right)$ in Eq.~\eqref{C_factor} is not affected by the backreaction. 
Thus the correction on the warp factor $f(\phi_\text{tot})$ in Eq.~\eqref{ssr:warp} can be expanded as $f(\bar{\phi}_\text{tot}(\tau)) = \left( 1 - 4 \frac{\bar{\phi}_\text{br}(\tau)}{\bar{\phi}_\text{bg}(\tau)} \right) f(\bar{\phi}_\text{bg}(\tau))$.
Using the above results, the sound speed with backreaction becomes
\begin{widetext}
\begin{equation}\label{cstot}
c_{s,\text{tot}}^2 =  c_{s,\text{bg}}^2 + \left(1-c_{s,\text{bg}}^2\right) \left[ 2 \Delta_H - 2 \left( \bar{\phi}_\mathrm{bg} - \frac{\bar{\phi}_\mathrm{bg}^2 }{\bar{\phi}_i} \right) \frac{ \dot{\Delta}_H -  \dot{\epsilon}_\text{br} }{\dot{\bar{\phi}}_\text{bg}} + \left( 2 - 4 \frac{\bar{\phi}_\text{bg}}{\bar{\phi}_i} \right) \epsilon_\text{br} \right]  ~.
\end{equation} 
\end{widetext}

\begin{figure}[h]
    \centering
    \includegraphics[width=0.45\textwidth]{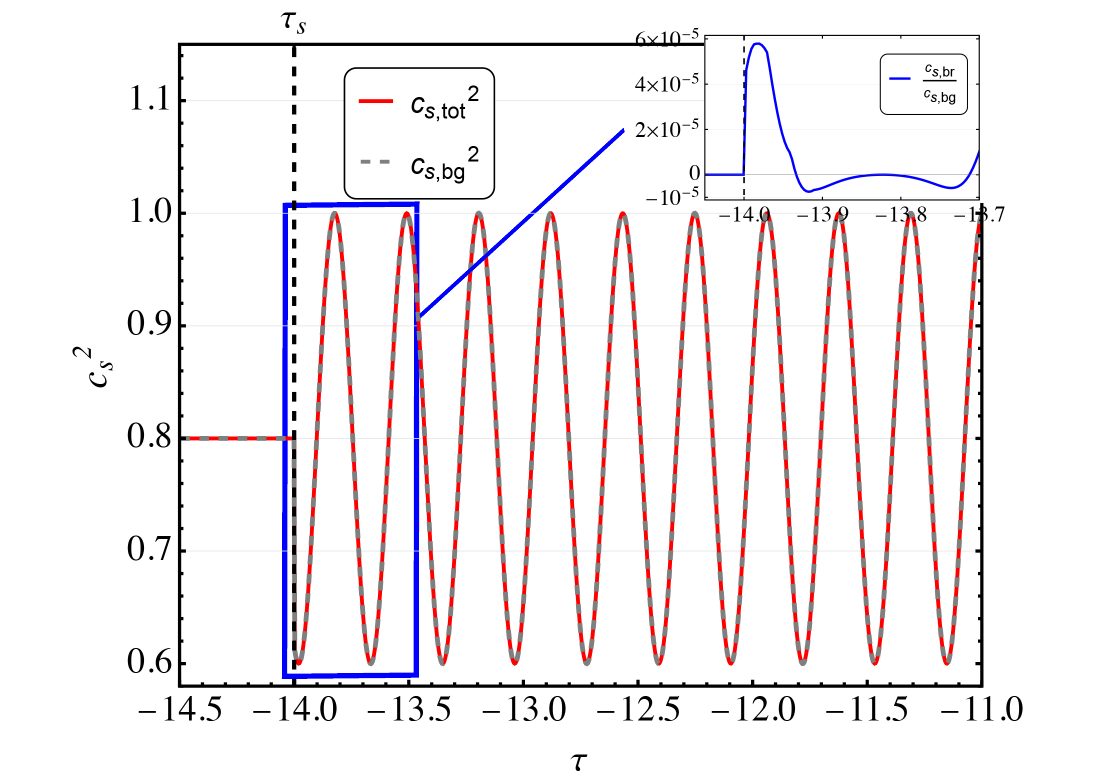}
    \caption{ 
    The numerical results of sound speeds $c_{s,\text{tot}}^2$ (red solid) and $c_{s,\text{bg}}^2$ (gray dashed), and their ratio is shown by the blue curve in the subplot.}
    \label{fig:cs}
\end{figure}

The evolutions of the sound speed $c_{s,\text{tot}/\text{bg}}^{2}$ are shown in Fig.~\ref{fig:cs}. The subplot shows the comparison with the correction manually amplified to show the difference more explicitly. 
It can be seen that the correction is large at the beginning of the oscillation stage and decays afterward, which is consistent with the behavior of $\epsilon_\text{br}$ shown in Fig.~\ref{fig:slowroll}. It is also worth noting that the backreaction does not cause the $c_s$ to become larger than untiy. 

\subsection{Backreaction effect on the curvature perturbations} \label{sec3b}

\begin{figure}[htbp]
    \centering
    \includegraphics[width=0.5\textwidth]{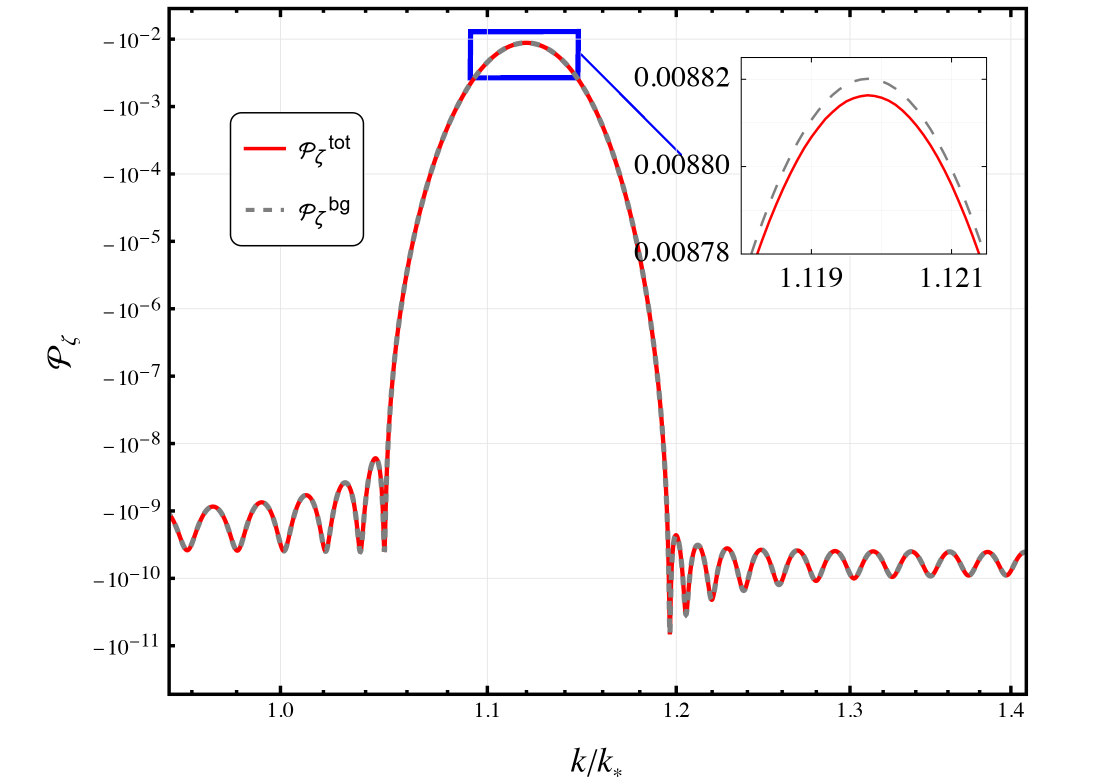}
    \caption{
    The numerical results of the curvature power spectra $\mathcal{P}^{\rm tot}_{\zeta}$ (red solid) and $\mathcal{P}^{\rm bg}_{\zeta}$ (gray dashed).
    The small subplot shows a zoom-in to the power spectra around the peak. 
    }
    \label{fig:pps}
\end{figure}

With the backreaction effects on the background quantities and the sound speed given in previous sections, we shall examine the backreaction on the power spectrum of curvature perturbations $\zeta$, which is given by
\be \label{BunchDaviesSpectrum}
\mathcal{P}_{\zeta} (k)
=\frac{ k^{3}}{2\pi^2} |\zeta_k |^2 ~,
\ee
and can be decomposed as $\mathcal{P}^{\rm tot}_{\zeta} = \mathcal{P}_{\zeta}^\text{bg} + \mathcal{P}_{\zeta}^\text{br}$.
The dynamics of $\zeta$ satisfies the Mukhanov-Sasaki equation \cite{Sasaki:1986hm,Mukhanov:1988jd}
\begin{equation}\label{MSeqq}
    \frac{\ddd^2 v_k}{\ddd \tau^2} + \l(c_{s,\text{tot}}^2 k^2 - \frac{1}{z} \frac{\ddd^2 z}{\ddd \tau^2}\r) v_k=0 ~,
\end{equation}
with  $v \equiv z \zeta$ and $ z = \sqrt{2 \epsilon_\text{tot}} a / c_{s,\text{tot}}$.
By taking the initial condition as the Bunch-Davies vacuum $v_k(\tau)=e^{- i \left( 1-2 \xi \right)^2 k \tau} / [ \sqrt{2 k} \left(1-2 \xi\right)^{1/4} ]$,
we numerically solved $\zeta$ at the horizon-crossing time $\tau=1/k$. The result for the power spectrum is shown in Fig.~\ref{fig:pps}.
We find that the peak with backreaction $\mathcal{P}^{\rm tot}_{\zeta} (k_{\rm peak})$ is smaller than that without backreaction $\mathcal{P}^{\rm bg}_{\zeta} (k_{\rm peak})$, where $k_\mathrm{peak}$ denotes the position of resonant peak. Subsequently, 
the relative correction $\left( \mathcal{P}^{\rm tot}_{\zeta}(k_{\rm peak})-\mathcal{P}^{\rm bg}_{\zeta}(k_{\rm peak}) \right)/\mathcal{P}^{\rm bg}_{\zeta}(k_{\rm peak}) \approx - 4.3 \times 10^{-4} $ is negative.
Therefore, the backreaction effect does not break the perturbativity of SSR mechanism.

\section{Conclusions}\label{Conclusion}

In this paper, we investigate the backreaction effect of the enhanced small-scale curvature perturbations in the SSR-DBI inflation. 
We find that the backreaction effects on both the background and perturbations can be neglected and do not spoil the previous results in the SSR-DBI inflation model in the absence of backreaction.

As a starting point, we review the SSR-DBI inflation model~\cite{Chen:2020uhe} and adopt the methods presented in Ref.~\cite{Gong:2022tfu} to calculate the backreaction effects of perturbations on the background. 
After integrating out the quadratic curvature perturbation action, we obtain the effective Lagrangian \eqref{br:brlag} which describes the backreaction effect on the inflationary background dynamics. 

Based on this effective Lagrangian, we numerically solve the evolution of the Hubble parameter $H$, the slow-roll parameter $\epsilon$, and the sound speed $c_s^2$ with backreaction.
We find that the backreaction effect causes a tiny correction to the Hubble parameter in an order of $10^{-7}$ compared to the original results in Ref.~\cite{Chen:2020uhe}, as shown in Fig.~\ref{fig:hubble}.
For the slow-roll parameter, as displayed in Fig.~\ref{fig:slowroll}, the correction resulting from the backreaction effect exhibits an oscillation that begins with an amplitude between $ -0.3$ and $0.1$, but gradually decreases to a value as small as $-2 \Delta_H \sim 10^{-7}$.
Additionally, the results presented in Fig.~\ref{fig:cs} tell us that the sound speed is slightly altered by curvature perturbations.
Applying the sound speed with/without backreaction to the Mukhanov-Sasaki equation \eqref{MSeqq}, we numerically solve the curvature perturbation $\zeta$ and plot the corresponding power spectra $\mathcal{P}_{\zeta}$ in Fig.~\ref{fig:pps}. 
We conclude that the backreaction effect on the amplification efficiency of curvature perturbations and the PBH formation predicted by the SSR mechanism is fairly small.

\acknowledgments

We would like to thank Yi-Fu Cai, Maria Mylova, Xin Ren, Chon Man Sou, Yi Wang, Xuan Ye and Yuhang Zhu for valuable suggestions and discussions. 
This work is supported in part by National Key R\&D Program of China (2021YFC2203100), by NSFC (11961131007, 12261131497, 12003029), by CAS young interdisciplinary innovation team (JCTD-2022-20), by 111 Project for ``Observational and Theoretical Research on Dark Matter and Dark Energy" (B23042), by Fundamental Research Funds for Central Universities, by CSC Innovation Talent Funds, by USTC Fellowship for International Cooperation, by USTC Research Funds of the Double First-Class Initiative, by CAS project for young scientists in basic research (YSBR-006). 
C.Chen is supported by the Jockey Club Institute for Advanced Study at HKUST.
We acknowledge the use of computing clusters {\it LINDA} and {\it JUDY} of the particle cosmology group at USTC.

\appendix

\begin{widetext}

\section{EFT description of SSR-DBI inflation}\label{app:EFT_SSR}

In this appendix, we shall give an EFT description for SSR-DBI inflation. EFT provides a generic description of metric perturbations around a quasi-de Sitter background \cite{Cheung:2007st}. Even without the knowledge of the full UV theory, EFT can remove the unnecessary high-energy degrees of freedom, while keeping track of their influence on the low-energy physics \cite{Weinberg:1978kz, Polchinski:1983gv, Polchinski:1992ed, Georgi:1993mps, Weinberg:2021exr, Burgess:2020tbq, Green:2022ovz}.
As the inflation vacuum is time-evolving, the general EFT action for inflation can be built by operators with broken-time translation symmetry \cite{Cheung:2007st, Weinberg:2008hq}.
For the single field inflation model $\phi(t, \vec{x})=\phi_0(t)+\delta \phi(t, \vec{x})$, one can work in the unitary gauge $\delta \phi(t, \vec{x}) = 0$ \cite{Cheung:2007st}. 
Besides the Riemann tensor $R_{\mu \nu \rho \sigma}$ that is already invariant under diffeomorphism transformation, operators satisfying this reduced symmetry in unitary gauge include tensors with upper index 0: $g^{00}$ and $R^{00}$, extrinsic curvature tensor $K_{\mu \nu} \equiv h_\mu{}^\sigma \nabla_\sigma n_\nu$ (where $n_\mu=\frac{\partial_\mu t}{\sqrt{-g^{\mu \nu} \partial_\mu t \partial_\nu t}}$ and $h_{\mu \nu} \equiv g_{\mu \nu}+n_\mu n_\nu$ are the normal vector and the induced metric of the hypersurface, respectively), arbitrary time-dependent function $f(t)$ and the covariant derivatives of the aforementioned operators. As $R^{00}$ can be written in terms of $g^{00}$ and $h^{\mu\nu}$, while $n^{\mu}$ can be written in terms of $g^{00}$ and $K_{\mu\nu}$, the most generic action is given by the following form \cite{Cheung:2007st},
\begin{equation}
S=\int d^4 x \sqrt{-g} F\left(R_{\mu \nu \rho \sigma}, g^{00}, K_{\mu \nu}, \nabla_\mu, t\right) ~,
\end{equation}
which can be expanded perturbatively as
\be \label{EFT_action}
\begin{aligned}
S_\text{EFT}
= &\int \ddd t \ddd ^3x \sqrt{-g}
\l[
\frac{M_{\mathrm{Pl}}^{2}}{2} R - c(t) g^{00} - \Lambda(t) 
+ { M_{2}(t)^{4} \over 2!} (g^{00}+1 )^{2} 
+ { M_{3}(t)^{4} \over 3!} (g^{00}+1)^{3} + \cdots \r.
\\
& - {\bar{M}_{1}(t)^{3} \over 2} (g^{00}+1) \delta K_{\mu}{}^{\mu}
- {\bar{M}_{2}(t)^{2} \over 2} (\delta K_{\mu}{}^{\mu})^{2} 
- {\bar{M}_{3}(t)^{2} \over 2} \delta K_{\nu}{}^{\mu} \delta K_{\mu}{}^{\nu} + \cdots
\\
&\l. + m_1^2 (t) h^{\mu \nu} \partial_\mu g^{00} \partial_\nu g^{00} + \lambda_1 (t) \delta R^2+\lambda_2 (t) \delta R_{\mu \nu} \delta R^{\mu \nu} + \mu_1^2 (t) \left( 1 + g^{00} \right) \delta R + \cdots \r] ~,
\end{aligned}
\ee
where $M_\text{Pl}$ is the reduced Planck mass, $R$ is the Ricci scalar,  $c(t)$, $\Lambda(t)$, $M_{2,3}(t)$, $\bar M_{1,2,3}(t)$, $\lambda_{1,2}(t)$, $\mu_1(t)$, $m_1 (t)$ are arbitrary time-dependent functions. The broken time diffeomorphism is restored by performing an infinitesimal transformation $t \rightarrow t + \pi(x)$, which will induce transformations for time-dependent functions in the action as
\begin{equation}
f(t) \rightarrow f\left(t+\pi\left(t, x^i\right)\right)=f(t)+\dot{f}(t) \pi\left(t, x^i\right)+\frac{1}{2} \ddot{f}(t) \pi\left(t, x^i\right)^2+\ldots ~,
\end{equation}
and requiring $\pi$ transform as $\pi(x) \rightarrow \widetilde{\pi}(\widetilde{x}(x))=\pi(x)-\xi^0(x)$ under the time translation $t \rightarrow \tilde{t}=t+\xi^0(x)$. 
This procedure is called the St{\" u}kelberg trick. The $\pi$ field that arises from the broken time diffeomorphism is the Goldstone boson, and is related to the curvature perturbation via relation $\zeta = - H \pi$ \cite{Cheung:2007st}. By invoking Arnowitt-Deser-Misner decomposition $d s^2=g_{\mu \nu} d x^\mu d x^\nu=-N^2 d t^2+h_{i j}\left(d x^i+N^i d t\right)\left(d x^j+N^j d t\right)$ and solving the constraints for Lagrangian multipliers: the lapse function $N$ and the shift functions $N^i$, one obtains a quadratic action describing the dynamics of $\pi$ \cite{Cheung:2007st, Gleyzes:2013ooa},
\be \label{pert_action_2}
S_{\pi}^{(2)}
= \int \ddd t \ddd ^3x a^{3}(t) 
\l[ \alpha (t) \dot{\pi}^{2} - \beta(t) { (\nabla\pi)^{2} \over a^{2}(t) } \r]  ~,
\ee
where the coefficients are defined as
\bl \label{pert_coeff}
\alpha(t) &= c(t) + 2 M_{2}(t)^{4} + \frac{3}{4} {\bar{M}_{1}(t)^{6} \over M_{\mathrm{Pl}}^{2}}  ~,
\\
\beta(t) &= c(t) - \frac{1}{4} {\bar{M}_{1}(t)^{6} \over M_{\mathrm{Pl}}^{2} } 
+ \frac{1}{2} \l( \dot{M}_{1}(t)^{3} + H \bar{M}_{1}(t)^{3} \r)  ~.
\el
We refer readers to \cite{Cabass:2022avo} for a recent review on EFT in cosmology.
The sound speed $c_s$ can be read directly from the action \eqref{pert_action_2},
\be \label{eft_cs}
c_{s}^{2}(t) = { \beta(t) \over \alpha(t) } ~.
\ee 
By matching with the DBI action, one can directly obtain the EFT coefficients as
\be\label{EFT_dictionary}
\begin{aligned}
c(t) &= \Lambda(t) = \frac12 \dot{\bar{\phi}}^2 \l( 1 - f(\bar{\phi}) \dot{\bar{\phi}}^2 \r)^{-1/2} ~,
\\
M_{n}(t)^{4} &= \frac14 f(\bar{\phi}) \dot{\bar{\phi}}^4 \l(1 - f(\bar{\phi}) \dot{\bar{\phi}}^2 \r)^{-3/2} ~,
\\
\bar M_1(t)^3 & = \lambda_{1,2}(t) = \mu_1^2(t) = m_1^2(t) =  0  ~,
\\
\bar M_2(t)^2 &= - \bar M_3(t)^2 = - M_{\mathrm{Pl}}^{2} ~,
\end{aligned}
\ee
and coefficients $\alpha(t)$ and $\beta(t)$ in Eq.~\eqref{pert_action_2} become
\be \label{alphabeta}
\begin{aligned}
\beta_\text{DBI}(t) &= \frac12 \dot{\bar{\phi}}^2 \l( 1 - f(\bar{\phi}) \dot{\bar{\phi}}^2 \r)^{-1/2} ~,
\\
\alpha_\text{DBI}(t) &= \frac12 \dot{\bar{\phi}}^2 \l( 1 - f(\bar{\phi}) \dot{\bar{\phi}}^2 \r)^{-3/2} ~.
\end{aligned}
\ee
with the above coefficients, Eq.~\eqref{eft_cs} also gives the sound speed in DBI inflation $c_s^2 = 1 - f(\bar{\phi}) \dot{\bar{\phi}}^2$.
Notice that the above EFT description works when $H$ is below the energy scale
\be \label{cutoff}
E^4 \sim 16 \pi^{2} M_{\mathrm{Pl}}^{2} |\dot{H}| \frac{c_{s}^{5}}{1-c_{s}^{2}} \sim 8 \pi^2 H^2 M_{\mathrm{Pl}}^2 \frac{\epsilon}{\xi} ~,
\ee
namely, the non-renormalizable interactions become insignificant~\cite{Cheung:2007st}.
In Eq.~\eqref{cutoff}, an approximation of time-averaged sound speed $c_s^2 = 1 - 2 \xi$ is adopted. For typical parameter choices in the SSR-DBI inflation, $H/M_{\rm Pl} \sim 10^{-5}$ and $\epsilon = 0.01$, the condition \eqref{cutoff} is satisfied.

\section{One-Loop effective action } \label{app:eff_action}

Here we derive the Lagrangian with backreaction effect from the one-loop effective action in the SSR-DBI model following the method presented by Ref.~\cite{Gong:2022tfu}, in which the backreaction is obtained by integrating out the high-order perturbations.
To get this effective action, we start from the partition function $Z[J]$ in Eq.~\eqref{action_1} with the perturbed inflation field as $\phi = \bar{\phi} + \phi_{\rm per}$, which can be perturbatively expanded around the classical background up to the second order as
\begin{equation}\label{app:partition_function}
Z[J] = \int \mathcal{D} \phi \exp\l\{ i \int \ddd ^{4}x \sqrt{-g}  \left( \mathcal{L}[\bar{\phi}] + J \bar{\phi} \right)
+ {1\over2} \int \sqrt{-g} \ddd^4 x \sqrt{-g} \ddd^4 y {\delta^2 \mathcal{L} [g, \phi] \over \delta\phi(x) \delta\phi(y)} \Big|_{\bar{\phi}} \phi_{\rm per}(x)  \phi_{\rm per}(y)  \r\} ~.
\end{equation}
The last term in the above formula represents the contribution from the perturbation $\phi_\text{per}$,
where the integral over  $\phi_\text{per}$ can be carried out by converting the quadratic term to a Gaussian integral \cite{Peskin:1995ev},
\be \label{app:gauss_int}
\begin{aligned}
\int\mathcal{D} \eta \exp\l(
{i\over2} \int \ddd^4 x \ddd^4 y \eta(x) \mathcal{A}(x,y) \eta(y) \r)
=
\text{const} \times \l( \det\l[ - i \mathcal{A}(x,y) \r] \r)^{-1/2} ~,
\end{aligned}
\ee
where $\mathcal{A}(x,y)$ is an operator, and the constant can be dropped by the normalization of vacuum state \cite{Peskin:1995ev}.
The effective action for $\bar{\phi}$ is \cite{Gong:2022tfu}
\be\label{effectActionDef}
\begin{aligned}
\Gamma [\bar{\phi}] 
= & - i \ln Z[J] - \int \ddd ^4 x \sqrt{-g} J \bar{\phi}
  ~.
\end{aligned}
\ee
Substituting Eqs.~\eqref{app:partition_function} and \eqref{app:gauss_int} into the above, and using the identity $\ln\l[\det(-i\mathcal{A}) \r]= \text{Tr}\l[ \ln(-i\mathcal{A}) \r]$ \cite{Peskin:1995ev}, Eq.~(\ref{effectActionDef}) becomes 
\be
\begin{aligned}
\Gamma [\bar{\phi}] 
=
\int \ddd ^{4}x \sqrt{-g}  \mathcal{L}[\bar{\phi}]  
- \frac{1}{2} \mathrm{Tr}\l[ \ln \l( - i \mathcal{A} \r) \r]  
\equiv
\int \ddd ^{4}x \sqrt{-g}  
\l(
\mathcal{L}_\text{bg}[\bar{\phi}] + \mathcal{L}_\text{br}[\bar{\phi}]
\r) ~,
\end{aligned}
\ee
for the SSR-DBI model, the background and the one-loop correction are
\begin{equation}\label{app:L_br}
   \mathcal{L}_\text{bg} = \mathcal{L}_\text{DBI} [\bar\phi] ~,
   \quad
   \mathcal{L}_\text{br}
   = - \frac{i}{2} \int {\ddd^4 k \over (2 \pi)^4} \ln \l( - i \mathcal{A} \r) ~,
\end{equation}
where $k$ denotes the physical momentum and we have taken the continuous limit of summation.
The explicit form of $\mathcal{A}$ can be obtained from  Eq.~\eqref{pert_action_2} using integration by parts,
\be \label{app:S2}
S^{(2)}
= \int \ddd \tau \ddd ^3 x \eta(\tau, \mathbf{x})
\l[ \frac{\ddd ^2}{\ddd  \tau^2} - c_s(\tau)^2 { \nabla^{2} \over a^{2}(\tau) } + \mathcal{M}^2 \r] \eta(\tau, \mathbf{x}) ~,
\ee
where $\eta \equiv \pi \l(a^2 \alpha\r)^{-1/2}$ is a normalized variable, the effective mass term is $\mathcal{M}^2 = \frac14 \l( 3 - 2 s \r)^2 H_\text{tot}^2 - H_\text{tot} \dot s$.
The operator $\mathcal{A}$ can be read by comparing Eq.~\eqref{app:S2} with Eq.~\eqref{app:gauss_int},
\be \label{A_E}
\mathcal{A}
= \frac{\ddd ^2}{\ddd  \tau^2} - c_s(\tau)^2 { \nabla^{2} \over a^{2}(\tau) } + \mathcal{M}^2 ~,
\ee
Substituting Eq.~\eqref{A_E} to Eq.~\eqref{app:L_br}, one gets
\begin{equation}
    \mathcal{L}_\text{br} =
    - \frac12 c_s^{-3} \l. \frac{\partial}{\partial\gamma } \l[ \int {\ddd ^4 \tilde k' \over (2 \pi)^4} {1 \over \l(\tilde k'{}^2 + \mathcal{M}^2 \r)^\gamma} \r] \r|_{\gamma = 0}  
    ~,
\end{equation}
where $k'{}^0 \equiv ik^0$ and $k'{}^i \equiv  c_s k^i$, and the identity 
$\int \frac{\ddd ^4 \tilde k }{(2 \pi)^4} \ln \left(\tilde k^2+\mathcal{M}^2\right)=\left. - \frac{\partial}{\partial\gamma } \l[ \int \frac{\ddd ^4 \tilde k }{(2 \pi)^4} \frac{1}{\left(\tilde k^2+\mathcal{M}^2\right)^\gamma} \r]\right|_{\gamma=0}$ is used \cite{Gong:2022tfu}. 
To prevent the divergence of the above integration, we introduce a cutoff at the sound horizon $\Lambda_k \equiv c_s^{-1} a H$ and only the modes of $|k'|<\Lambda_k$ contribute to the integration,
since the resonant mode of curvature perturbation in SSR-DBI inflation reaches its maximum at the sound horizon and the backreaction effect shall be relatively small for $|k'|>\Lambda_k$. 
So we carry out the integration for $|k'|<\Lambda_k$ and obtain Eq.~\eqref{br:brlag}.

\end{widetext}

\bibliographystyle{apsrev4-1}
\bibliography{main}

\begin{thebibliography}{91}%
\makeatletter
\providecommand \@ifxundefined [1]{%
 \@ifx{#1\undefined}
}%
\providecommand \@ifnum [1]{%
 \ifnum #1\expandafter \@firstoftwo
 \else \expandafter \@secondoftwo
 \fi
}%
\providecommand \@ifx [1]{%
 \ifx #1\expandafter \@firstoftwo
 \else \expandafter \@secondoftwo
 \fi
}%
\providecommand \natexlab [1]{#1}%
\providecommand \enquote  [1]{``#1''}%
\providecommand \bibnamefont  [1]{#1}%
\providecommand \bibfnamefont [1]{#1}%
\providecommand \citenamefont [1]{#1}%
\providecommand \href@noop [0]{\@secondoftwo}%
\providecommand \href [0]{\begingroup \@sanitize@url \@href}%
\providecommand \@href[1]{\@@startlink{#1}\@@href}%
\providecommand \@@href[1]{\endgroup#1\@@endlink}%
\providecommand \@sanitize@url [0]{\catcode `\\12\catcode `\$12\catcode
  `\&12\catcode `\#12\catcode `\^12\catcode `\_12\catcode `\%12\relax}%
\providecommand \@@startlink[1]{}%
\providecommand \@@endlink[0]{}%
\providecommand \url  [0]{\begingroup\@sanitize@url \@url }%
\providecommand \@url [1]{\endgroup\@href {#1}{\urlprefix }}%
\providecommand \urlprefix  [0]{URL }%
\providecommand \Eprint [0]{\href }%
\providecommand \doibase [0]{http://dx.doi.org/}%
\providecommand \selectlanguage [0]{\@gobble}%
\providecommand \bibinfo  [0]{\@secondoftwo}%
\providecommand \bibfield  [0]{\@secondoftwo}%
\providecommand \translation [1]{[#1]}%
\providecommand \BibitemOpen [0]{}%
\providecommand \bibitemStop [0]{}%
\providecommand \bibitemNoStop [0]{.\EOS\space}%
\providecommand \EOS [0]{\spacefactor3000\relax}%
\providecommand \BibitemShut  [1]{\csname bibitem#1\endcsname}%
\let\auto@bib@innerbib\@empty
\bibitem [{\citenamefont {Brill}\ and\ \citenamefont
  {Hartle}(1964)}]{Brill:1964zz}%
  \BibitemOpen
  \bibfield  {author} {\bibinfo {author} {\bibfnamefont {D.~R.}\ \bibnamefont
  {Brill}}\ and\ \bibinfo {author} {\bibfnamefont {J.~B.}\ \bibnamefont
  {Hartle}},\ }\href {\doibase 10.1103/PhysRev.135.B271} {\bibfield  {journal}
  {\bibinfo  {journal} {Phys. Rev.}\ }\textbf {\bibinfo {volume} {135}},\
  \bibinfo {pages} {B271} (\bibinfo {year} {1964})}\BibitemShut {NoStop}%
\bibitem [{\citenamefont {Isaacson}(1968{\natexlab{a}})}]{Isaacson:1968hbi}%
  \BibitemOpen
  \bibfield  {author} {\bibinfo {author} {\bibfnamefont {R.~A.}\ \bibnamefont
  {Isaacson}},\ }\href {\doibase 10.1103/PhysRev.166.1263} {\bibfield
  {journal} {\bibinfo  {journal} {Phys. Rev.}\ }\textbf {\bibinfo {volume}
  {166}},\ \bibinfo {pages} {1263} (\bibinfo {year}
  {1968}{\natexlab{a}})}\BibitemShut {NoStop}%
\bibitem [{\citenamefont {Isaacson}(1968{\natexlab{b}})}]{Isaacson:1968zza}%
  \BibitemOpen
  \bibfield  {author} {\bibinfo {author} {\bibfnamefont {R.~A.}\ \bibnamefont
  {Isaacson}},\ }\href {\doibase 10.1103/PhysRev.166.1272} {\bibfield
  {journal} {\bibinfo  {journal} {Phys. Rev.}\ }\textbf {\bibinfo {volume}
  {166}},\ \bibinfo {pages} {1272} (\bibinfo {year}
  {1968}{\natexlab{b}})}\BibitemShut {NoStop}%
\bibitem [{\citenamefont {Abramo}\ \emph {et~al.}(1997)\citenamefont {Abramo},
  \citenamefont {Brandenberger},\ and\ \citenamefont
  {Mukhanov}}]{Abramo:1997hu}%
  \BibitemOpen
  \bibfield  {author} {\bibinfo {author} {\bibfnamefont {L.~R.~W.}\
  \bibnamefont {Abramo}}, \bibinfo {author} {\bibfnamefont {R.~H.}\
  \bibnamefont {Brandenberger}}, \ and\ \bibinfo {author} {\bibfnamefont
  {V.~F.}\ \bibnamefont {Mukhanov}},\ }\href {\doibase
  10.1103/PhysRevD.56.3248} {\bibfield  {journal} {\bibinfo  {journal} {Phys.
  Rev. D}\ }\textbf {\bibinfo {volume} {56}},\ \bibinfo {pages} {3248}
  (\bibinfo {year} {1997})},\ \Eprint {http://arxiv.org/abs/gr-qc/9704037}
  {arXiv:gr-qc/9704037} \BibitemShut {NoStop}%
\bibitem [{\citenamefont {Giovannini}(2020)}]{Giovannini:2019oii}%
  \BibitemOpen
  \bibfield  {author} {\bibinfo {author} {\bibfnamefont {M.}~\bibnamefont
  {Giovannini}},\ }\href {\doibase 10.1016/j.ppnp.2020.103774} {\bibfield
  {journal} {\bibinfo  {journal} {Prog. Part. Nucl. Phys.}\ }\textbf {\bibinfo
  {volume} {112}},\ \bibinfo {pages} {103774} (\bibinfo {year} {2020})},\
  \Eprint {http://arxiv.org/abs/1912.07065} {arXiv:1912.07065 [astro-ph.CO]}
  \BibitemShut {NoStop}%
\bibitem [{\citenamefont {Adamek}\ \emph {et~al.}(2019)\citenamefont {Adamek},
  \citenamefont {Clarkson}, \citenamefont {Daverio}, \citenamefont {Durrer},\
  and\ \citenamefont {Kunz}}]{Adamek:2017mzb}%
  \BibitemOpen
  \bibfield  {author} {\bibinfo {author} {\bibfnamefont {J.}~\bibnamefont
  {Adamek}}, \bibinfo {author} {\bibfnamefont {C.}~\bibnamefont {Clarkson}},
  \bibinfo {author} {\bibfnamefont {D.}~\bibnamefont {Daverio}}, \bibinfo
  {author} {\bibfnamefont {R.}~\bibnamefont {Durrer}}, \ and\ \bibinfo {author}
  {\bibfnamefont {M.}~\bibnamefont {Kunz}},\ }\href {\doibase
  10.1088/1361-6382/aaeca5} {\bibfield  {journal} {\bibinfo  {journal} {Class.
  Quant. Grav.}\ }\textbf {\bibinfo {volume} {36}},\ \bibinfo {pages} {014001}
  (\bibinfo {year} {2019})},\ \Eprint {http://arxiv.org/abs/1706.09309}
  {arXiv:1706.09309 [astro-ph.CO]} \BibitemShut {NoStop}%
\bibitem [{\citenamefont {Schander}\ and\ \citenamefont
  {Thiemann}(2021)}]{Schander:2021pgt}%
  \BibitemOpen
  \bibfield  {author} {\bibinfo {author} {\bibfnamefont {S.}~\bibnamefont
  {Schander}}\ and\ \bibinfo {author} {\bibfnamefont {T.}~\bibnamefont
  {Thiemann}},\ }\href {\doibase 10.3389/fspas.2021.692198} {\bibfield
  {journal} {\bibinfo  {journal} {Front. Astron. Space Sci.}\ }\textbf
  {\bibinfo {volume} {0}},\ \bibinfo {pages} {113} (\bibinfo {year} {2021})},\
  \Eprint {http://arxiv.org/abs/2106.06043} {arXiv:2106.06043 [gr-qc]}
  \BibitemShut {NoStop}%
\bibitem [{\citenamefont {Hu}\ and\ \citenamefont
  {Verdaguer}(2020)}]{Hu:2020luk}%
  \BibitemOpen
  \bibfield  {author} {\bibinfo {author} {\bibfnamefont {B.-L.~B.}\
  \bibnamefont {Hu}}\ and\ \bibinfo {author} {\bibfnamefont {E.}~\bibnamefont
  {Verdaguer}},\ }\href {\doibase 10.1017/9780511667497} {\emph {\bibinfo
  {title} {{Semiclassical and Stochastic Gravity}: {Quantum Field Effects on
  Curved Spacetime}}}},\ Cambridge Monographs on Mathematical Physics\
  (\bibinfo  {publisher} {Cambridge University Press},\ \bibinfo {address}
  {Cambridge},\ \bibinfo {year} {2020})\BibitemShut {NoStop}%
\bibitem [{\citenamefont {Sasaki}\ \emph {et~al.}(2018)\citenamefont {Sasaki},
  \citenamefont {Suyama}, \citenamefont {Tanaka},\ and\ \citenamefont
  {Yokoyama}}]{Sasaki:2018dmp}%
  \BibitemOpen
  \bibfield  {author} {\bibinfo {author} {\bibfnamefont {M.}~\bibnamefont
  {Sasaki}}, \bibinfo {author} {\bibfnamefont {T.}~\bibnamefont {Suyama}},
  \bibinfo {author} {\bibfnamefont {T.}~\bibnamefont {Tanaka}}, \ and\ \bibinfo
  {author} {\bibfnamefont {S.}~\bibnamefont {Yokoyama}},\ }\href {\doibase
  10.1088/1361-6382/aaa7b4} {\bibfield  {journal} {\bibinfo  {journal} {Class.
  Quant. Grav.}\ }\textbf {\bibinfo {volume} {35}},\ \bibinfo {pages} {063001}
  (\bibinfo {year} {2018})},\ \Eprint {http://arxiv.org/abs/1801.05235}
  {arXiv:1801.05235 [astro-ph.CO]} \BibitemShut {NoStop}%
\bibitem [{\citenamefont {Carr}\ and\ \citenamefont
  {Kuhnel}(2020)}]{Carr:2020xqk}%
  \BibitemOpen
  \bibfield  {author} {\bibinfo {author} {\bibfnamefont {B.}~\bibnamefont
  {Carr}}\ and\ \bibinfo {author} {\bibfnamefont {F.}~\bibnamefont {Kuhnel}},\
  }\href {\doibase 10.1146/annurev-nucl-050520-125911} {\bibfield  {journal}
  {\bibinfo  {journal} {Ann. Rev. Nucl. Part. Sci.}\ }\textbf {\bibinfo
  {volume} {70}},\ \bibinfo {pages} {355} (\bibinfo {year} {2020})},\ \Eprint
  {http://arxiv.org/abs/2006.02838} {arXiv:2006.02838 [astro-ph.CO]}
  \BibitemShut {NoStop}%
\bibitem [{\citenamefont {Escriv\`a}\ \emph {et~al.}(2022)\citenamefont
  {Escriv\`a}, \citenamefont {Kuhnel},\ and\ \citenamefont
  {Tada}}]{Escriva:2022duf}%
  \BibitemOpen
  \bibfield  {author} {\bibinfo {author} {\bibfnamefont {A.}~\bibnamefont
  {Escriv\`a}}, \bibinfo {author} {\bibfnamefont {F.}~\bibnamefont {Kuhnel}}, \
  and\ \bibinfo {author} {\bibfnamefont {Y.}~\bibnamefont {Tada}},\ }\href@noop
  {} {\  (\bibinfo {year} {2022})},\ \Eprint {http://arxiv.org/abs/2211.05767}
  {arXiv:2211.05767 [astro-ph.CO]} \BibitemShut {NoStop}%
\bibitem [{\citenamefont {Garcia-Bellido}\ and\ \citenamefont
  {Ruiz~Morales}(2017)}]{Garcia-Bellido:2017mdw}%
  \BibitemOpen
  \bibfield  {author} {\bibinfo {author} {\bibfnamefont {J.}~\bibnamefont
  {Garcia-Bellido}}\ and\ \bibinfo {author} {\bibfnamefont {E.}~\bibnamefont
  {Ruiz~Morales}},\ }\href {\doibase 10.1016/j.dark.2017.09.007} {\bibfield
  {journal} {\bibinfo  {journal} {Phys. Dark Univ.}\ }\textbf {\bibinfo
  {volume} {18}},\ \bibinfo {pages} {47} (\bibinfo {year} {2017})},\ \Eprint
  {http://arxiv.org/abs/1702.03901} {arXiv:1702.03901 [astro-ph.CO]}
  \BibitemShut {NoStop}%
\bibitem [{\citenamefont {Germani}\ and\ \citenamefont
  {Prokopec}(2017)}]{Germani:2017bcs}%
  \BibitemOpen
  \bibfield  {author} {\bibinfo {author} {\bibfnamefont {C.}~\bibnamefont
  {Germani}}\ and\ \bibinfo {author} {\bibfnamefont {T.}~\bibnamefont
  {Prokopec}},\ }\href {\doibase 10.1016/j.dark.2017.09.001} {\bibfield
  {journal} {\bibinfo  {journal} {Phys. Dark Univ.}\ }\textbf {\bibinfo
  {volume} {18}},\ \bibinfo {pages} {6} (\bibinfo {year} {2017})},\ \Eprint
  {http://arxiv.org/abs/1706.04226} {arXiv:1706.04226 [astro-ph.CO]}
  \BibitemShut {NoStop}%
\bibitem [{\citenamefont {Motohashi}\ and\ \citenamefont
  {Hu}(2017)}]{Motohashi:2017kbs}%
  \BibitemOpen
  \bibfield  {author} {\bibinfo {author} {\bibfnamefont {H.}~\bibnamefont
  {Motohashi}}\ and\ \bibinfo {author} {\bibfnamefont {W.}~\bibnamefont {Hu}},\
  }\href {\doibase 10.1103/PhysRevD.96.063503} {\bibfield  {journal} {\bibinfo
  {journal} {Phys. Rev. D}\ }\textbf {\bibinfo {volume} {96}},\ \bibinfo
  {pages} {063503} (\bibinfo {year} {2017})},\ \Eprint
  {http://arxiv.org/abs/1706.06784} {arXiv:1706.06784 [astro-ph.CO]}
  \BibitemShut {NoStop}%
\bibitem [{\citenamefont {Pattison}\ \emph {et~al.}(2017)\citenamefont
  {Pattison}, \citenamefont {Vennin}, \citenamefont {Assadullahi},\ and\
  \citenamefont {Wands}}]{Pattison:2017mbe}%
  \BibitemOpen
  \bibfield  {author} {\bibinfo {author} {\bibfnamefont {C.}~\bibnamefont
  {Pattison}}, \bibinfo {author} {\bibfnamefont {V.}~\bibnamefont {Vennin}},
  \bibinfo {author} {\bibfnamefont {H.}~\bibnamefont {Assadullahi}}, \ and\
  \bibinfo {author} {\bibfnamefont {D.}~\bibnamefont {Wands}},\ }\href
  {\doibase 10.1088/1475-7516/2017/10/046} {\bibfield  {journal} {\bibinfo
  {journal} {JCAP}\ }\textbf {\bibinfo {volume} {10}},\ \bibinfo {pages} {046}
  (\bibinfo {year} {2017})},\ \Eprint {http://arxiv.org/abs/1707.00537}
  {arXiv:1707.00537 [hep-th]} \BibitemShut {NoStop}%
\bibitem [{\citenamefont {Di}\ and\ \citenamefont {Gong}(2018)}]{Di:2017ndc}%
  \BibitemOpen
  \bibfield  {author} {\bibinfo {author} {\bibfnamefont {H.}~\bibnamefont
  {Di}}\ and\ \bibinfo {author} {\bibfnamefont {Y.}~\bibnamefont {Gong}},\
  }\href {\doibase 10.1088/1475-7516/2018/07/007} {\bibfield  {journal}
  {\bibinfo  {journal} {JCAP}\ }\textbf {\bibinfo {volume} {07}},\ \bibinfo
  {pages} {007} (\bibinfo {year} {2018})},\ \Eprint
  {http://arxiv.org/abs/1707.09578} {arXiv:1707.09578 [astro-ph.CO]}
  \BibitemShut {NoStop}%
\bibitem [{\citenamefont {Byrnes}\ \emph {et~al.}(2019)\citenamefont {Byrnes},
  \citenamefont {Cole},\ and\ \citenamefont {Patil}}]{Byrnes:2018txb}%
  \BibitemOpen
  \bibfield  {author} {\bibinfo {author} {\bibfnamefont {C.~T.}\ \bibnamefont
  {Byrnes}}, \bibinfo {author} {\bibfnamefont {P.~S.}\ \bibnamefont {Cole}}, \
  and\ \bibinfo {author} {\bibfnamefont {S.~P.}\ \bibnamefont {Patil}},\ }\href
  {\doibase 10.1088/1475-7516/2019/06/028} {\bibfield  {journal} {\bibinfo
  {journal} {JCAP}\ }\textbf {\bibinfo {volume} {06}},\ \bibinfo {pages} {028}
  (\bibinfo {year} {2019})},\ \Eprint {http://arxiv.org/abs/1811.11158}
  {arXiv:1811.11158 [astro-ph.CO]} \BibitemShut {NoStop}%
\bibitem [{\citenamefont {Passaglia}\ \emph {et~al.}(2019)\citenamefont
  {Passaglia}, \citenamefont {Hu},\ and\ \citenamefont
  {Motohashi}}]{Passaglia:2018ixg}%
  \BibitemOpen
  \bibfield  {author} {\bibinfo {author} {\bibfnamefont {S.}~\bibnamefont
  {Passaglia}}, \bibinfo {author} {\bibfnamefont {W.}~\bibnamefont {Hu}}, \
  and\ \bibinfo {author} {\bibfnamefont {H.}~\bibnamefont {Motohashi}},\ }\href
  {\doibase 10.1103/PhysRevD.99.043536} {\bibfield  {journal} {\bibinfo
  {journal} {Phys. Rev. D}\ }\textbf {\bibinfo {volume} {99}},\ \bibinfo
  {pages} {043536} (\bibinfo {year} {2019})},\ \Eprint
  {http://arxiv.org/abs/1812.08243} {arXiv:1812.08243 [astro-ph.CO]}
  \BibitemShut {NoStop}%
\bibitem [{\citenamefont {Xu}\ \emph {et~al.}(2020)\citenamefont {Xu},
  \citenamefont {Liu}, \citenamefont {Gao},\ and\ \citenamefont
  {Guo}}]{Xu:2019bdp}%
  \BibitemOpen
  \bibfield  {author} {\bibinfo {author} {\bibfnamefont {W.-T.}\ \bibnamefont
  {Xu}}, \bibinfo {author} {\bibfnamefont {J.}~\bibnamefont {Liu}}, \bibinfo
  {author} {\bibfnamefont {T.-J.}\ \bibnamefont {Gao}}, \ and\ \bibinfo
  {author} {\bibfnamefont {Z.-K.}\ \bibnamefont {Guo}},\ }\href {\doibase
  10.1103/PhysRevD.101.023505} {\bibfield  {journal} {\bibinfo  {journal}
  {Phys. Rev. D}\ }\textbf {\bibinfo {volume} {101}},\ \bibinfo {pages}
  {023505} (\bibinfo {year} {2020})},\ \Eprint
  {http://arxiv.org/abs/1907.05213} {arXiv:1907.05213 [astro-ph.CO]}
  \BibitemShut {NoStop}%
\bibitem [{\citenamefont {Tasinato}(2021)}]{Tasinato:2020vdk}%
  \BibitemOpen
  \bibfield  {author} {\bibinfo {author} {\bibfnamefont {G.}~\bibnamefont
  {Tasinato}},\ }\href {\doibase 10.1103/PhysRevD.103.023535} {\bibfield
  {journal} {\bibinfo  {journal} {Phys. Rev. D}\ }\textbf {\bibinfo {volume}
  {103}},\ \bibinfo {pages} {023535} (\bibinfo {year} {2021})},\ \Eprint
  {http://arxiv.org/abs/2012.02518} {arXiv:2012.02518 [hep-th]} \BibitemShut
  {NoStop}%
\bibitem [{\citenamefont {Fu}\ \emph {et~al.}(2020)\citenamefont {Fu},
  \citenamefont {Wu},\ and\ \citenamefont {Yu}}]{Fu:2020lob}%
  \BibitemOpen
  \bibfield  {author} {\bibinfo {author} {\bibfnamefont {C.}~\bibnamefont
  {Fu}}, \bibinfo {author} {\bibfnamefont {P.}~\bibnamefont {Wu}}, \ and\
  \bibinfo {author} {\bibfnamefont {H.}~\bibnamefont {Yu}},\ }\href {\doibase
  10.1103/PhysRevD.102.043527} {\bibfield  {journal} {\bibinfo  {journal}
  {Phys. Rev. D}\ }\textbf {\bibinfo {volume} {102}},\ \bibinfo {pages}
  {043527} (\bibinfo {year} {2020})},\ \Eprint
  {http://arxiv.org/abs/2006.03768} {arXiv:2006.03768 [astro-ph.CO]}
  \BibitemShut {NoStop}%
\bibitem [{\citenamefont {Inomata}\ \emph
  {et~al.}(2022{\natexlab{a}})\citenamefont {Inomata}, \citenamefont
  {McDonough},\ and\ \citenamefont {Hu}}]{Inomata:2021tpx}%
  \BibitemOpen
  \bibfield  {author} {\bibinfo {author} {\bibfnamefont {K.}~\bibnamefont
  {Inomata}}, \bibinfo {author} {\bibfnamefont {E.}~\bibnamefont {McDonough}},
  \ and\ \bibinfo {author} {\bibfnamefont {W.}~\bibnamefont {Hu}},\ }\href
  {\doibase 10.1088/1475-7516/2022/02/031} {\bibfield  {journal} {\bibinfo
  {journal} {JCAP}\ }\textbf {\bibinfo {volume} {02}},\ \bibinfo {pages} {031}
  (\bibinfo {year} {2022}{\natexlab{a}})},\ \Eprint
  {http://arxiv.org/abs/2110.14641} {arXiv:2110.14641 [astro-ph.CO]}
  \BibitemShut {NoStop}%
\bibitem [{\citenamefont {Kawai}\ and\ \citenamefont
  {Kim}(2021)}]{Kawai:2021edk}%
  \BibitemOpen
  \bibfield  {author} {\bibinfo {author} {\bibfnamefont {S.}~\bibnamefont
  {Kawai}}\ and\ \bibinfo {author} {\bibfnamefont {J.}~\bibnamefont {Kim}},\
  }\href {\doibase 10.1103/PhysRevD.104.083545} {\bibfield  {journal} {\bibinfo
   {journal} {Phys. Rev. D}\ }\textbf {\bibinfo {volume} {104}},\ \bibinfo
  {pages} {083545} (\bibinfo {year} {2021})},\ \Eprint
  {http://arxiv.org/abs/2108.01340} {arXiv:2108.01340 [astro-ph.CO]}
  \BibitemShut {NoStop}%
\bibitem [{\citenamefont {Zheng}\ \emph {et~al.}(2022)\citenamefont {Zheng},
  \citenamefont {Shi},\ and\ \citenamefont {Qiu}}]{ZhengRuiFeng:2021zoz}%
  \BibitemOpen
  \bibfield  {author} {\bibinfo {author} {\bibfnamefont {R.}~\bibnamefont
  {Zheng}}, \bibinfo {author} {\bibfnamefont {J.}~\bibnamefont {Shi}}, \ and\
  \bibinfo {author} {\bibfnamefont {T.}~\bibnamefont {Qiu}},\ }\href {\doibase
  10.1088/1674-1137/ac42bd} {\bibfield  {journal} {\bibinfo  {journal} {Chin.
  Phys. C}\ }\textbf {\bibinfo {volume} {46}},\ \bibinfo {pages} {045103}
  (\bibinfo {year} {2022})},\ \Eprint {http://arxiv.org/abs/2106.04303}
  {arXiv:2106.04303 [astro-ph.CO]} \BibitemShut {NoStop}%
\bibitem [{\citenamefont {Balaji}\ \emph {et~al.}(2022)\citenamefont {Balaji},
  \citenamefont {Silk},\ and\ \citenamefont {Wu}}]{Balaji:2022rsy}%
  \BibitemOpen
  \bibfield  {author} {\bibinfo {author} {\bibfnamefont {S.}~\bibnamefont
  {Balaji}}, \bibinfo {author} {\bibfnamefont {J.}~\bibnamefont {Silk}}, \ and\
  \bibinfo {author} {\bibfnamefont {Y.-P.}\ \bibnamefont {Wu}},\ }\href
  {\doibase 10.1088/1475-7516/2022/06/008} {\bibfield  {journal} {\bibinfo
  {journal} {JCAP}\ }\textbf {\bibinfo {volume} {06}},\ \bibinfo {pages} {008}
  (\bibinfo {year} {2022})},\ \Eprint {http://arxiv.org/abs/2202.00700}
  {arXiv:2202.00700 [astro-ph.CO]} \BibitemShut {NoStop}%
\bibitem [{\citenamefont {Qiu}\ \emph {et~al.}(2023)\citenamefont {Qiu},
  \citenamefont {Wang},\ and\ \citenamefont {Zheng}}]{Qiu:2022klm}%
  \BibitemOpen
  \bibfield  {author} {\bibinfo {author} {\bibfnamefont {T.}~\bibnamefont
  {Qiu}}, \bibinfo {author} {\bibfnamefont {W.}~\bibnamefont {Wang}}, \ and\
  \bibinfo {author} {\bibfnamefont {R.}~\bibnamefont {Zheng}},\ }\href
  {\doibase 10.1103/PhysRevD.107.083018} {\bibfield  {journal} {\bibinfo
  {journal} {Phys. Rev. D}\ }\textbf {\bibinfo {volume} {107}},\ \bibinfo
  {pages} {083018} (\bibinfo {year} {2023})},\ \Eprint
  {http://arxiv.org/abs/2212.03403} {arXiv:2212.03403 [astro-ph.CO]}
  \BibitemShut {NoStop}%
\bibitem [{\citenamefont {Fu}\ and\ \citenamefont {Chen}(2023)}]{Fu:2022ssq}%
  \BibitemOpen
  \bibfield  {author} {\bibinfo {author} {\bibfnamefont {C.}~\bibnamefont
  {Fu}}\ and\ \bibinfo {author} {\bibfnamefont {C.}~\bibnamefont {Chen}},\
  }\href {\doibase 10.1088/1475-7516/2023/05/005} {\bibfield  {journal}
  {\bibinfo  {journal} {JCAP}\ }\textbf {\bibinfo {volume} {05}},\ \bibinfo
  {pages} {005} (\bibinfo {year} {2023})},\ \Eprint
  {http://arxiv.org/abs/2211.11387} {arXiv:2211.11387 [astro-ph.CO]}
  \BibitemShut {NoStop}%
\bibitem [{\citenamefont {Kasai}\ \emph {et~al.}(2022)\citenamefont {Kasai},
  \citenamefont {Kawasaki},\ and\ \citenamefont {Murai}}]{Kasai:2022vhq}%
  \BibitemOpen
  \bibfield  {author} {\bibinfo {author} {\bibfnamefont {K.}~\bibnamefont
  {Kasai}}, \bibinfo {author} {\bibfnamefont {M.}~\bibnamefont {Kawasaki}}, \
  and\ \bibinfo {author} {\bibfnamefont {K.}~\bibnamefont {Murai}},\ }\href
  {\doibase 10.1088/1475-7516/2022/10/048} {\bibfield  {journal} {\bibinfo
  {journal} {JCAP}\ }\textbf {\bibinfo {volume} {10}},\ \bibinfo {pages} {048}
  (\bibinfo {year} {2022})},\ \Eprint {http://arxiv.org/abs/2205.10148}
  {arXiv:2205.10148 [astro-ph.CO]} \BibitemShut {NoStop}%
\bibitem [{\citenamefont {Fu}\ and\ \citenamefont {Wang}(2023)}]{Fu:2022ypp}%
  \BibitemOpen
  \bibfield  {author} {\bibinfo {author} {\bibfnamefont {C.}~\bibnamefont
  {Fu}}\ and\ \bibinfo {author} {\bibfnamefont {S.-J.}\ \bibnamefont {Wang}},\
  }\href {\doibase 10.1088/1475-7516/2023/06/012} {\bibfield  {journal}
  {\bibinfo  {journal} {JCAP}\ }\textbf {\bibinfo {volume} {06}},\ \bibinfo
  {pages} {012} (\bibinfo {year} {2023})},\ \Eprint
  {http://arxiv.org/abs/2211.03523} {arXiv:2211.03523 [astro-ph.CO]}
  \BibitemShut {NoStop}%
\bibitem [{\citenamefont {Pi}\ and\ \citenamefont {Wang}(2023)}]{Pi:2022zxs}%
  \BibitemOpen
  \bibfield  {author} {\bibinfo {author} {\bibfnamefont {S.}~\bibnamefont
  {Pi}}\ and\ \bibinfo {author} {\bibfnamefont {J.}~\bibnamefont {Wang}},\
  }\href {\doibase 10.1088/1475-7516/2023/06/018} {\bibfield  {journal}
  {\bibinfo  {journal} {JCAP}\ }\textbf {\bibinfo {volume} {06}},\ \bibinfo
  {pages} {018} (\bibinfo {year} {2023})},\ \Eprint
  {http://arxiv.org/abs/2209.14183} {arXiv:2209.14183 [astro-ph.CO]}
  \BibitemShut {NoStop}%
\bibitem [{\citenamefont {Cai}\ \emph {et~al.}(2023)\citenamefont {Cai},
  \citenamefont {Zhu},\ and\ \citenamefont {Piao}}]{Cai:2023uhc}%
  \BibitemOpen
  \bibfield  {author} {\bibinfo {author} {\bibfnamefont {Y.}~\bibnamefont
  {Cai}}, \bibinfo {author} {\bibfnamefont {M.}~\bibnamefont {Zhu}}, \ and\
  \bibinfo {author} {\bibfnamefont {Y.-S.}\ \bibnamefont {Piao}},\ }\href@noop
  {} {\  (\bibinfo {year} {2023})},\ \Eprint {http://arxiv.org/abs/2305.10933}
  {arXiv:2305.10933 [gr-qc]} \BibitemShut {NoStop}%
\bibitem [{\citenamefont {Cai}\ \emph {et~al.}(2018)\citenamefont {Cai},
  \citenamefont {Tong}, \citenamefont {Wang},\ and\ \citenamefont
  {Yan}}]{Cai:2018tuh}%
  \BibitemOpen
  \bibfield  {author} {\bibinfo {author} {\bibfnamefont {Y.-F.}\ \bibnamefont
  {Cai}}, \bibinfo {author} {\bibfnamefont {X.}~\bibnamefont {Tong}}, \bibinfo
  {author} {\bibfnamefont {D.-G.}\ \bibnamefont {Wang}}, \ and\ \bibinfo
  {author} {\bibfnamefont {S.-F.}\ \bibnamefont {Yan}},\ }\href {\doibase
  10.1103/PhysRevLett.121.081306} {\bibfield  {journal} {\bibinfo  {journal}
  {Phys. Rev. Lett.}\ }\textbf {\bibinfo {volume} {121}},\ \bibinfo {pages}
  {081306} (\bibinfo {year} {2018})},\ \Eprint
  {http://arxiv.org/abs/1805.03639} {arXiv:1805.03639 [astro-ph.CO]}
  \BibitemShut {NoStop}%
\bibitem [{\citenamefont {Chen}\ and\ \citenamefont
  {Cai}(2019)}]{Chen:2019zza}%
  \BibitemOpen
  \bibfield  {author} {\bibinfo {author} {\bibfnamefont {C.}~\bibnamefont
  {Chen}}\ and\ \bibinfo {author} {\bibfnamefont {Y.-F.}\ \bibnamefont {Cai}},\
  }\href {\doibase 10.1088/1475-7516/2019/10/068} {\bibfield  {journal}
  {\bibinfo  {journal} {JCAP}\ }\textbf {\bibinfo {volume} {10}},\ \bibinfo
  {pages} {068} (\bibinfo {year} {2019})},\ \Eprint
  {http://arxiv.org/abs/1908.03942} {arXiv:1908.03942 [astro-ph.CO]}
  \BibitemShut {NoStop}%
\bibitem [{\citenamefont {Cai}\ \emph {et~al.}(2020)\citenamefont {Cai},
  \citenamefont {Guo}, \citenamefont {Liu}, \citenamefont {Liu},\ and\
  \citenamefont {Yang}}]{Cai:2019bmk}%
  \BibitemOpen
  \bibfield  {author} {\bibinfo {author} {\bibfnamefont {R.-G.}\ \bibnamefont
  {Cai}}, \bibinfo {author} {\bibfnamefont {Z.-K.}\ \bibnamefont {Guo}},
  \bibinfo {author} {\bibfnamefont {J.}~\bibnamefont {Liu}}, \bibinfo {author}
  {\bibfnamefont {L.}~\bibnamefont {Liu}}, \ and\ \bibinfo {author}
  {\bibfnamefont {X.-Y.}\ \bibnamefont {Yang}},\ }\href {\doibase
  10.1088/1475-7516/2020/06/013} {\bibfield  {journal} {\bibinfo  {journal}
  {JCAP}\ }\textbf {\bibinfo {volume} {06}},\ \bibinfo {pages} {013} (\bibinfo
  {year} {2020})},\ \Eprint {http://arxiv.org/abs/1912.10437} {arXiv:1912.10437
  [astro-ph.CO]} \BibitemShut {NoStop}%
\bibitem [{\citenamefont {Ashoorioon}\ \emph {et~al.}(2021)\citenamefont
  {Ashoorioon}, \citenamefont {Rostami},\ and\ \citenamefont
  {Firouzjaee}}]{Ashoorioon:2019xqc}%
  \BibitemOpen
  \bibfield  {author} {\bibinfo {author} {\bibfnamefont {A.}~\bibnamefont
  {Ashoorioon}}, \bibinfo {author} {\bibfnamefont {A.}~\bibnamefont {Rostami}},
  \ and\ \bibinfo {author} {\bibfnamefont {J.~T.}\ \bibnamefont {Firouzjaee}},\
  }\href {\doibase 10.1007/JHEP07(2021)087} {\bibfield  {journal} {\bibinfo
  {journal} {JHEP}\ }\textbf {\bibinfo {volume} {07}},\ \bibinfo {pages} {087}
  (\bibinfo {year} {2021})},\ \Eprint {http://arxiv.org/abs/1912.13326}
  {arXiv:1912.13326 [astro-ph.CO]} \BibitemShut {NoStop}%
\bibitem [{\citenamefont {Zhou}\ \emph {et~al.}(2020)\citenamefont {Zhou},
  \citenamefont {Jiang}, \citenamefont {Cai}, \citenamefont {Sasaki},\ and\
  \citenamefont {Pi}}]{Zhou:2020kkf}%
  \BibitemOpen
  \bibfield  {author} {\bibinfo {author} {\bibfnamefont {Z.}~\bibnamefont
  {Zhou}}, \bibinfo {author} {\bibfnamefont {J.}~\bibnamefont {Jiang}},
  \bibinfo {author} {\bibfnamefont {Y.-F.}\ \bibnamefont {Cai}}, \bibinfo
  {author} {\bibfnamefont {M.}~\bibnamefont {Sasaki}}, \ and\ \bibinfo {author}
  {\bibfnamefont {S.}~\bibnamefont {Pi}},\ }\href {\doibase
  10.1103/PhysRevD.102.103527} {\bibfield  {journal} {\bibinfo  {journal}
  {Phys. Rev. D}\ }\textbf {\bibinfo {volume} {102}},\ \bibinfo {pages}
  {103527} (\bibinfo {year} {2020})},\ \Eprint
  {http://arxiv.org/abs/2010.03537} {arXiv:2010.03537 [astro-ph.CO]}
  \BibitemShut {NoStop}%
\bibitem [{\citenamefont {Chen}\ \emph {et~al.}(2020)\citenamefont {Chen},
  \citenamefont {Ma},\ and\ \citenamefont {Cai}}]{Chen:2020uhe}%
  \BibitemOpen
  \bibfield  {author} {\bibinfo {author} {\bibfnamefont {C.}~\bibnamefont
  {Chen}}, \bibinfo {author} {\bibfnamefont {X.-H.}\ \bibnamefont {Ma}}, \ and\
  \bibinfo {author} {\bibfnamefont {Y.-F.}\ \bibnamefont {Cai}},\ }\href
  {\doibase 10.1103/PhysRevD.102.063526} {\bibfield  {journal} {\bibinfo
  {journal} {Phys. Rev. D}\ }\textbf {\bibinfo {volume} {102}},\ \bibinfo
  {pages} {063526} (\bibinfo {year} {2020})},\ \Eprint
  {http://arxiv.org/abs/2003.03821} {arXiv:2003.03821 [astro-ph.CO]}
  \BibitemShut {NoStop}%
\bibitem [{\citenamefont {Cai}\ \emph {et~al.}(2021{\natexlab{a}})\citenamefont
  {Cai}, \citenamefont {Chen},\ and\ \citenamefont {Fu}}]{Cai:2021wzd}%
  \BibitemOpen
  \bibfield  {author} {\bibinfo {author} {\bibfnamefont {R.-G.}\ \bibnamefont
  {Cai}}, \bibinfo {author} {\bibfnamefont {C.}~\bibnamefont {Chen}}, \ and\
  \bibinfo {author} {\bibfnamefont {C.}~\bibnamefont {Fu}},\ }\href {\doibase
  10.1103/PhysRevD.104.083537} {\bibfield  {journal} {\bibinfo  {journal}
  {Phys. Rev. D}\ }\textbf {\bibinfo {volume} {104}},\ \bibinfo {pages}
  {083537} (\bibinfo {year} {2021}{\natexlab{a}})},\ \Eprint
  {http://arxiv.org/abs/2108.03422} {arXiv:2108.03422 [astro-ph.CO]}
  \BibitemShut {NoStop}%
\bibitem [{\citenamefont {Linde}\ \emph {et~al.}(2013)\citenamefont {Linde},
  \citenamefont {Mooij},\ and\ \citenamefont {Pajer}}]{Linde:2012bt}%
  \BibitemOpen
  \bibfield  {author} {\bibinfo {author} {\bibfnamefont {A.}~\bibnamefont
  {Linde}}, \bibinfo {author} {\bibfnamefont {S.}~\bibnamefont {Mooij}}, \ and\
  \bibinfo {author} {\bibfnamefont {E.}~\bibnamefont {Pajer}},\ }\href
  {\doibase 10.1103/PhysRevD.87.103506} {\bibfield  {journal} {\bibinfo
  {journal} {Phys. Rev. D}\ }\textbf {\bibinfo {volume} {87}},\ \bibinfo
  {pages} {103506} (\bibinfo {year} {2013})},\ \Eprint
  {http://arxiv.org/abs/1212.1693} {arXiv:1212.1693 [hep-th]} \BibitemShut
  {NoStop}%
\bibitem [{\citenamefont {Braglia}\ \emph {et~al.}(2020)\citenamefont
  {Braglia}, \citenamefont {Hazra}, \citenamefont {Finelli}, \citenamefont
  {Smoot}, \citenamefont {Sriramkumar},\ and\ \citenamefont
  {Starobinsky}}]{Braglia:2020eai}%
  \BibitemOpen
  \bibfield  {author} {\bibinfo {author} {\bibfnamefont {M.}~\bibnamefont
  {Braglia}}, \bibinfo {author} {\bibfnamefont {D.~K.}\ \bibnamefont {Hazra}},
  \bibinfo {author} {\bibfnamefont {F.}~\bibnamefont {Finelli}}, \bibinfo
  {author} {\bibfnamefont {G.~F.}\ \bibnamefont {Smoot}}, \bibinfo {author}
  {\bibfnamefont {L.}~\bibnamefont {Sriramkumar}}, \ and\ \bibinfo {author}
  {\bibfnamefont {A.~A.}\ \bibnamefont {Starobinsky}},\ }\href {\doibase
  10.1088/1475-7516/2020/08/001} {\bibfield  {journal} {\bibinfo  {journal}
  {JCAP}\ }\textbf {\bibinfo {volume} {08}},\ \bibinfo {pages} {001} (\bibinfo
  {year} {2020})},\ \Eprint {http://arxiv.org/abs/2005.02895} {arXiv:2005.02895
  [astro-ph.CO]} \BibitemShut {NoStop}%
\bibitem [{\citenamefont {Liu}(2023)}]{Liu:2021rgq}%
  \BibitemOpen
  \bibfield  {author} {\bibinfo {author} {\bibfnamefont {L.-H.}\ \bibnamefont
  {Liu}},\ }\href {\doibase 10.1088/1674-1137/ac9d28} {\bibfield  {journal}
  {\bibinfo  {journal} {Chin. Phys. C}\ }\textbf {\bibinfo {volume} {47}},\
  \bibinfo {pages} {1} (\bibinfo {year} {2023})},\ \Eprint
  {http://arxiv.org/abs/2107.07310} {arXiv:2107.07310 [astro-ph.CO]}
  \BibitemShut {NoStop}%
\bibitem [{\citenamefont {Meng}\ \emph {et~al.}(2023)\citenamefont {Meng},
  \citenamefont {Yuan},\ and\ \citenamefont {Huang}}]{Meng:2022low}%
  \BibitemOpen
  \bibfield  {author} {\bibinfo {author} {\bibfnamefont {D.-S.}\ \bibnamefont
  {Meng}}, \bibinfo {author} {\bibfnamefont {C.}~\bibnamefont {Yuan}}, \ and\
  \bibinfo {author} {\bibfnamefont {Q.-G.}\ \bibnamefont {Huang}},\ }\href
  {\doibase 10.1007/s11433-022-2095-5} {\bibfield  {journal} {\bibinfo
  {journal} {Sci. China Phys. Mech. Astron.}\ }\textbf {\bibinfo {volume}
  {66}},\ \bibinfo {pages} {280411} (\bibinfo {year} {2023})},\ \Eprint
  {http://arxiv.org/abs/2212.03577} {arXiv:2212.03577 [astro-ph.CO]}
  \BibitemShut {NoStop}%
\bibitem [{\citenamefont {Chen}\ \emph
  {et~al.}(2023{\natexlab{a}})\citenamefont {Chen}, \citenamefont {Ghoshal},
  \citenamefont {Lalak}, \citenamefont {Luo},\ and\ \citenamefont
  {Naskar}}]{Chen:2023lou}%
  \BibitemOpen
  \bibfield  {author} {\bibinfo {author} {\bibfnamefont {C.}~\bibnamefont
  {Chen}}, \bibinfo {author} {\bibfnamefont {A.}~\bibnamefont {Ghoshal}},
  \bibinfo {author} {\bibfnamefont {Z.}~\bibnamefont {Lalak}}, \bibinfo
  {author} {\bibfnamefont {Y.}~\bibnamefont {Luo}}, \ and\ \bibinfo {author}
  {\bibfnamefont {A.}~\bibnamefont {Naskar}},\ }\href@noop {} {\  (\bibinfo
  {year} {2023}{\natexlab{a}})},\ \Eprint {http://arxiv.org/abs/2305.12325}
  {arXiv:2305.12325 [astro-ph.CO]} \BibitemShut {NoStop}%
\bibitem [{\citenamefont {Ferrante}\ \emph {et~al.}(2023)\citenamefont
  {Ferrante}, \citenamefont {Franciolini}, \citenamefont {Iovino},\ and\
  \citenamefont {Urbano}}]{Ferrante:2023bgz}%
  \BibitemOpen
  \bibfield  {author} {\bibinfo {author} {\bibfnamefont {G.}~\bibnamefont
  {Ferrante}}, \bibinfo {author} {\bibfnamefont {G.}~\bibnamefont
  {Franciolini}}, \bibinfo {author} {\bibfnamefont {A.}~\bibnamefont {Iovino},
  \bibfnamefont {Junior.}}, \ and\ \bibinfo {author} {\bibfnamefont
  {A.}~\bibnamefont {Urbano}},\ }\href@noop {} {\  (\bibinfo {year} {2023})},\
  \Eprint {http://arxiv.org/abs/2305.13382} {arXiv:2305.13382 [astro-ph.CO]}
  \BibitemShut {NoStop}%
\bibitem [{\citenamefont {Ge}\ \emph {et~al.}(2023)\citenamefont {Ge},
  \citenamefont {Guo},\ and\ \citenamefont {Liu}}]{Ge:2023rrq}%
  \BibitemOpen
  \bibfield  {author} {\bibinfo {author} {\bibfnamefont {S.}~\bibnamefont
  {Ge}}, \bibinfo {author} {\bibfnamefont {J.}~\bibnamefont {Guo}}, \ and\
  \bibinfo {author} {\bibfnamefont {J.}~\bibnamefont {Liu}},\ }\href@noop {} {\
   (\bibinfo {year} {2023})},\ \Eprint {http://arxiv.org/abs/2309.01739}
  {arXiv:2309.01739 [hep-ph]} \BibitemShut {NoStop}%
\bibitem [{\citenamefont {Ezquiaga}\ \emph {et~al.}(2020)\citenamefont
  {Ezquiaga}, \citenamefont {Garc\'\i{}a-Bellido},\ and\ \citenamefont
  {Vennin}}]{Ezquiaga:2019ftu}%
  \BibitemOpen
  \bibfield  {author} {\bibinfo {author} {\bibfnamefont {J.~M.}\ \bibnamefont
  {Ezquiaga}}, \bibinfo {author} {\bibfnamefont {J.}~\bibnamefont
  {Garc\'\i{}a-Bellido}}, \ and\ \bibinfo {author} {\bibfnamefont
  {V.}~\bibnamefont {Vennin}},\ }\href {\doibase 10.1088/1475-7516/2020/03/029}
  {\bibfield  {journal} {\bibinfo  {journal} {JCAP}\ }\textbf {\bibinfo
  {volume} {03}},\ \bibinfo {pages} {029} (\bibinfo {year} {2020})},\ \Eprint
  {http://arxiv.org/abs/1912.05399} {arXiv:1912.05399 [astro-ph.CO]}
  \BibitemShut {NoStop}%
\bibitem [{\citenamefont {Atal}\ \emph {et~al.}(2020)\citenamefont {Atal},
  \citenamefont {Cid}, \citenamefont {Escriv\`a},\ and\ \citenamefont
  {Garriga}}]{Atal:2019erb}%
  \BibitemOpen
  \bibfield  {author} {\bibinfo {author} {\bibfnamefont {V.}~\bibnamefont
  {Atal}}, \bibinfo {author} {\bibfnamefont {J.}~\bibnamefont {Cid}}, \bibinfo
  {author} {\bibfnamefont {A.}~\bibnamefont {Escriv\`a}}, \ and\ \bibinfo
  {author} {\bibfnamefont {J.}~\bibnamefont {Garriga}},\ }\href {\doibase
  10.1088/1475-7516/2020/05/022} {\bibfield  {journal} {\bibinfo  {journal}
  {JCAP}\ }\textbf {\bibinfo {volume} {05}},\ \bibinfo {pages} {022} (\bibinfo
  {year} {2020})},\ \Eprint {http://arxiv.org/abs/1908.11357} {arXiv:1908.11357
  [astro-ph.CO]} \BibitemShut {NoStop}%
\bibitem [{\citenamefont {Figueroa}\ \emph {et~al.}(2021)\citenamefont
  {Figueroa}, \citenamefont {Raatikainen}, \citenamefont {Rasanen},\ and\
  \citenamefont {Tomberg}}]{Figueroa:2020jkf}%
  \BibitemOpen
  \bibfield  {author} {\bibinfo {author} {\bibfnamefont {D.~G.}\ \bibnamefont
  {Figueroa}}, \bibinfo {author} {\bibfnamefont {S.}~\bibnamefont
  {Raatikainen}}, \bibinfo {author} {\bibfnamefont {S.}~\bibnamefont
  {Rasanen}}, \ and\ \bibinfo {author} {\bibfnamefont {E.}~\bibnamefont
  {Tomberg}},\ }\href {\doibase 10.1103/PhysRevLett.127.101302} {\bibfield
  {journal} {\bibinfo  {journal} {Phys. Rev. Lett.}\ }\textbf {\bibinfo
  {volume} {127}},\ \bibinfo {pages} {101302} (\bibinfo {year} {2021})},\
  \Eprint {http://arxiv.org/abs/2012.06551} {arXiv:2012.06551 [astro-ph.CO]}
  \BibitemShut {NoStop}%
\bibitem [{\citenamefont {Cai}\ \emph {et~al.}(2021{\natexlab{b}})\citenamefont
  {Cai}, \citenamefont {Ma}, \citenamefont {Sasaki}, \citenamefont {Wang},\
  and\ \citenamefont {Zhou}}]{Cai:2021zsp}%
  \BibitemOpen
  \bibfield  {author} {\bibinfo {author} {\bibfnamefont {Y.-F.}\ \bibnamefont
  {Cai}}, \bibinfo {author} {\bibfnamefont {X.-H.}\ \bibnamefont {Ma}},
  \bibinfo {author} {\bibfnamefont {M.}~\bibnamefont {Sasaki}}, \bibinfo
  {author} {\bibfnamefont {D.-G.}\ \bibnamefont {Wang}}, \ and\ \bibinfo
  {author} {\bibfnamefont {Z.}~\bibnamefont {Zhou}},\ }\href@noop {} {\
  (\bibinfo {year} {2021}{\natexlab{b}})},\ \Eprint
  {http://arxiv.org/abs/2112.13836} {arXiv:2112.13836 [astro-ph.CO]}
  \BibitemShut {NoStop}%
\bibitem [{\citenamefont {Cai}\ \emph {et~al.}(2022)\citenamefont {Cai},
  \citenamefont {Ma}, \citenamefont {Sasaki}, \citenamefont {Wang},\ and\
  \citenamefont {Zhou}}]{Cai:2022erk}%
  \BibitemOpen
  \bibfield  {author} {\bibinfo {author} {\bibfnamefont {Y.-F.}\ \bibnamefont
  {Cai}}, \bibinfo {author} {\bibfnamefont {X.-H.}\ \bibnamefont {Ma}},
  \bibinfo {author} {\bibfnamefont {M.}~\bibnamefont {Sasaki}}, \bibinfo
  {author} {\bibfnamefont {D.-G.}\ \bibnamefont {Wang}}, \ and\ \bibinfo
  {author} {\bibfnamefont {Z.}~\bibnamefont {Zhou}},\ }\href@noop {} {\
  (\bibinfo {year} {2022})},\ \Eprint {http://arxiv.org/abs/2207.11910}
  {arXiv:2207.11910 [astro-ph.CO]} \BibitemShut {NoStop}%
\bibitem [{\citenamefont {Matsubara}\ and\ \citenamefont
  {Sasaki}(2022)}]{Matsubara:2022nbr}%
  \BibitemOpen
  \bibfield  {author} {\bibinfo {author} {\bibfnamefont {T.}~\bibnamefont
  {Matsubara}}\ and\ \bibinfo {author} {\bibfnamefont {M.}~\bibnamefont
  {Sasaki}},\ }\href {\doibase 10.1088/1475-7516/2022/10/094} {\bibfield
  {journal} {\bibinfo  {journal} {JCAP}\ }\textbf {\bibinfo {volume} {10}},\
  \bibinfo {pages} {094} (\bibinfo {year} {2022})},\ \Eprint
  {http://arxiv.org/abs/2208.02941} {arXiv:2208.02941 [astro-ph.CO]}
  \BibitemShut {NoStop}%
\bibitem [{\citenamefont {Gow}\ \emph {et~al.}(2022)\citenamefont {Gow},
  \citenamefont {Assadullahi}, \citenamefont {Jackson}, \citenamefont {Koyama},
  \citenamefont {Vennin},\ and\ \citenamefont {Wands}}]{Gow:2022jfb}%
  \BibitemOpen
  \bibfield  {author} {\bibinfo {author} {\bibfnamefont {A.~D.}\ \bibnamefont
  {Gow}}, \bibinfo {author} {\bibfnamefont {H.}~\bibnamefont {Assadullahi}},
  \bibinfo {author} {\bibfnamefont {J.~H.~P.}\ \bibnamefont {Jackson}},
  \bibinfo {author} {\bibfnamefont {K.}~\bibnamefont {Koyama}}, \bibinfo
  {author} {\bibfnamefont {V.}~\bibnamefont {Vennin}}, \ and\ \bibinfo {author}
  {\bibfnamefont {D.}~\bibnamefont {Wands}},\ }\href@noop {} {\  (\bibinfo
  {year} {2022})},\ \Eprint {http://arxiv.org/abs/2211.08348} {arXiv:2211.08348
  [astro-ph.CO]} \BibitemShut {NoStop}%
\bibitem [{\citenamefont {Pi}\ and\ \citenamefont {Sasaki}(2023)}]{Pi:2022ysn}%
  \BibitemOpen
  \bibfield  {author} {\bibinfo {author} {\bibfnamefont {S.}~\bibnamefont
  {Pi}}\ and\ \bibinfo {author} {\bibfnamefont {M.}~\bibnamefont {Sasaki}},\
  }\href {\doibase 10.1103/PhysRevLett.131.011002} {\bibfield  {journal}
  {\bibinfo  {journal} {Phys. Rev. Lett.}\ }\textbf {\bibinfo {volume} {131}},\
  \bibinfo {pages} {011002} (\bibinfo {year} {2023})},\ \Eprint
  {http://arxiv.org/abs/2211.13932} {arXiv:2211.13932 [astro-ph.CO]}
  \BibitemShut {NoStop}%
\bibitem [{\citenamefont {Cheng}\ \emph {et~al.}(2022)\citenamefont {Cheng},
  \citenamefont {Lee},\ and\ \citenamefont {Ng}}]{Cheng:2021lif}%
  \BibitemOpen
  \bibfield  {author} {\bibinfo {author} {\bibfnamefont {S.-L.}\ \bibnamefont
  {Cheng}}, \bibinfo {author} {\bibfnamefont {D.-S.}\ \bibnamefont {Lee}}, \
  and\ \bibinfo {author} {\bibfnamefont {K.-W.}\ \bibnamefont {Ng}},\ }\href
  {\doibase 10.1016/j.physletb.2022.136956} {\bibfield  {journal} {\bibinfo
  {journal} {Phys. Lett. B}\ }\textbf {\bibinfo {volume} {827}},\ \bibinfo
  {pages} {136956} (\bibinfo {year} {2022})},\ \Eprint
  {http://arxiv.org/abs/2106.09275} {arXiv:2106.09275 [astro-ph.CO]}
  \BibitemShut {NoStop}%
\bibitem [{\citenamefont {Caravano}\ \emph {et~al.}(2022)\citenamefont
  {Caravano}, \citenamefont {Komatsu}, \citenamefont {Lozanov},\ and\
  \citenamefont {Weller}}]{Caravano:2022epk}%
  \BibitemOpen
  \bibfield  {author} {\bibinfo {author} {\bibfnamefont {A.}~\bibnamefont
  {Caravano}}, \bibinfo {author} {\bibfnamefont {E.}~\bibnamefont {Komatsu}},
  \bibinfo {author} {\bibfnamefont {K.~D.}\ \bibnamefont {Lozanov}}, \ and\
  \bibinfo {author} {\bibfnamefont {J.}~\bibnamefont {Weller}},\ }\href@noop {}
  {\  (\bibinfo {year} {2022})},\ \Eprint {http://arxiv.org/abs/2204.12874}
  {arXiv:2204.12874 [astro-ph.CO]} \BibitemShut {NoStop}%
\bibitem [{\citenamefont {Yu}\ \emph {et~al.}(2023)\citenamefont {Yu},
  \citenamefont {Fu},\ and\ \citenamefont {Guo}}]{Yu:2023ity}%
  \BibitemOpen
  \bibfield  {author} {\bibinfo {author} {\bibfnamefont {Z.}~\bibnamefont
  {Yu}}, \bibinfo {author} {\bibfnamefont {C.}~\bibnamefont {Fu}}, \ and\
  \bibinfo {author} {\bibfnamefont {Z.-K.}\ \bibnamefont {Guo}},\ }\href@noop
  {} {\  (\bibinfo {year} {2023})},\ \Eprint {http://arxiv.org/abs/2307.03120}
  {arXiv:2307.03120 [gr-qc]} \BibitemShut {NoStop}%
\bibitem [{\citenamefont {Inomata}\ \emph
  {et~al.}(2022{\natexlab{b}})\citenamefont {Inomata}, \citenamefont
  {Braglia},\ and\ \citenamefont {Chen}}]{Inomata:2022yte}%
  \BibitemOpen
  \bibfield  {author} {\bibinfo {author} {\bibfnamefont {K.}~\bibnamefont
  {Inomata}}, \bibinfo {author} {\bibfnamefont {M.}~\bibnamefont {Braglia}}, \
  and\ \bibinfo {author} {\bibfnamefont {X.}~\bibnamefont {Chen}},\ }\href@noop
  {} {\  (\bibinfo {year} {2022}{\natexlab{b}})},\ \Eprint
  {http://arxiv.org/abs/2211.02586} {arXiv:2211.02586 [astro-ph.CO]}
  \BibitemShut {NoStop}%
\bibitem [{\citenamefont {Kristiano}\ and\ \citenamefont
  {Yokoyama}(2022)}]{Kristiano:2022maq}%
  \BibitemOpen
  \bibfield  {author} {\bibinfo {author} {\bibfnamefont {J.}~\bibnamefont
  {Kristiano}}\ and\ \bibinfo {author} {\bibfnamefont {J.}~\bibnamefont
  {Yokoyama}},\ }\href@noop {} {\  (\bibinfo {year} {2022})},\ \Eprint
  {http://arxiv.org/abs/2211.03395} {arXiv:2211.03395 [hep-th]} \BibitemShut
  {NoStop}%
\bibitem [{\citenamefont {Riotto}(2023)}]{Riotto:2023hoz}%
  \BibitemOpen
  \bibfield  {author} {\bibinfo {author} {\bibfnamefont {A.}~\bibnamefont
  {Riotto}},\ }\href@noop {} {\  (\bibinfo {year} {2023})},\ \Eprint
  {http://arxiv.org/abs/2301.00599} {arXiv:2301.00599 [astro-ph.CO]}
  \BibitemShut {NoStop}%
\bibitem [{\citenamefont {Choudhury}\ \emph
  {et~al.}(2023{\natexlab{a}})\citenamefont {Choudhury}, \citenamefont
  {Panda},\ and\ \citenamefont {Sami}}]{Choudhury:2023rks}%
  \BibitemOpen
  \bibfield  {author} {\bibinfo {author} {\bibfnamefont {S.}~\bibnamefont
  {Choudhury}}, \bibinfo {author} {\bibfnamefont {S.}~\bibnamefont {Panda}}, \
  and\ \bibinfo {author} {\bibfnamefont {M.}~\bibnamefont {Sami}},\ }\href@noop
  {} {\  (\bibinfo {year} {2023}{\natexlab{a}})},\ \Eprint
  {http://arxiv.org/abs/2303.06066} {arXiv:2303.06066 [astro-ph.CO]}
  \BibitemShut {NoStop}%
\bibitem [{\citenamefont {Choudhury}\ \emph
  {et~al.}(2023{\natexlab{b}})\citenamefont {Choudhury}, \citenamefont
  {Gangopadhyay},\ and\ \citenamefont {Sami}}]{Choudhury:2023vuj}%
  \BibitemOpen
  \bibfield  {author} {\bibinfo {author} {\bibfnamefont {S.}~\bibnamefont
  {Choudhury}}, \bibinfo {author} {\bibfnamefont {M.~R.}\ \bibnamefont
  {Gangopadhyay}}, \ and\ \bibinfo {author} {\bibfnamefont {M.}~\bibnamefont
  {Sami}},\ }\href@noop {} {\  (\bibinfo {year} {2023}{\natexlab{b}})},\
  \Eprint {http://arxiv.org/abs/2301.10000} {arXiv:2301.10000 [astro-ph.CO]}
  \BibitemShut {NoStop}%
\bibitem [{\citenamefont {Firouzjahi}(2023)}]{Firouzjahi:2023btw}%
  \BibitemOpen
  \bibfield  {author} {\bibinfo {author} {\bibfnamefont {H.}~\bibnamefont
  {Firouzjahi}},\ }\href@noop {} {\  (\bibinfo {year} {2023})},\ \Eprint
  {http://arxiv.org/abs/2305.01527} {arXiv:2305.01527 [astro-ph.CO]}
  \BibitemShut {NoStop}%
\bibitem [{\citenamefont {Franciolini}\ \emph {et~al.}(2023)\citenamefont
  {Franciolini}, \citenamefont {Iovino}, \citenamefont {Taoso},\ and\
  \citenamefont {Urbano}}]{Franciolini:2023lgy}%
  \BibitemOpen
  \bibfield  {author} {\bibinfo {author} {\bibfnamefont {G.}~\bibnamefont
  {Franciolini}}, \bibinfo {author} {\bibfnamefont {A.}~\bibnamefont {Iovino},
  \bibfnamefont {Junior.}}, \bibinfo {author} {\bibfnamefont {M.}~\bibnamefont
  {Taoso}}, \ and\ \bibinfo {author} {\bibfnamefont {A.}~\bibnamefont
  {Urbano}},\ }\href@noop {} {\  (\bibinfo {year} {2023})},\ \Eprint
  {http://arxiv.org/abs/2305.03491} {arXiv:2305.03491 [astro-ph.CO]}
  \BibitemShut {NoStop}%
\bibitem [{\citenamefont {Cheng}\ \emph {et~al.}(2023)\citenamefont {Cheng},
  \citenamefont {Lee},\ and\ \citenamefont {Ng}}]{Cheng:2023ikq}%
  \BibitemOpen
  \bibfield  {author} {\bibinfo {author} {\bibfnamefont {S.-L.}\ \bibnamefont
  {Cheng}}, \bibinfo {author} {\bibfnamefont {D.-S.}\ \bibnamefont {Lee}}, \
  and\ \bibinfo {author} {\bibfnamefont {K.-W.}\ \bibnamefont {Ng}},\
  }\href@noop {} {\  (\bibinfo {year} {2023})},\ \Eprint
  {http://arxiv.org/abs/2305.16810} {arXiv:2305.16810 [astro-ph.CO]}
  \BibitemShut {NoStop}%
\bibitem [{\citenamefont {Fumagalli}(2023)}]{Fumagalli:2023hpa}%
  \BibitemOpen
  \bibfield  {author} {\bibinfo {author} {\bibfnamefont {J.}~\bibnamefont
  {Fumagalli}},\ }\href@noop {} {\  (\bibinfo {year} {2023})},\ \Eprint
  {http://arxiv.org/abs/2305.19263} {arXiv:2305.19263 [astro-ph.CO]}
  \BibitemShut {NoStop}%
\bibitem [{\citenamefont {Chen}\ \emph
  {et~al.}(2023{\natexlab{b}})\citenamefont {Chen}, \citenamefont {Ota},
  \citenamefont {Zhu},\ and\ \citenamefont {Zhu}}]{Chen:2022dah}%
  \BibitemOpen
  \bibfield  {author} {\bibinfo {author} {\bibfnamefont {C.}~\bibnamefont
  {Chen}}, \bibinfo {author} {\bibfnamefont {A.}~\bibnamefont {Ota}}, \bibinfo
  {author} {\bibfnamefont {H.-Y.}\ \bibnamefont {Zhu}}, \ and\ \bibinfo
  {author} {\bibfnamefont {Y.}~\bibnamefont {Zhu}},\ }\href {\doibase
  10.1103/PhysRevD.107.083518} {\bibfield  {journal} {\bibinfo  {journal}
  {Phys. Rev. D}\ }\textbf {\bibinfo {volume} {107}},\ \bibinfo {pages}
  {083518} (\bibinfo {year} {2023}{\natexlab{b}})},\ \Eprint
  {http://arxiv.org/abs/2210.17176} {arXiv:2210.17176 [astro-ph.CO]}
  \BibitemShut {NoStop}%
\bibitem [{\citenamefont {Ota}\ \emph {et~al.}(2022)\citenamefont {Ota},
  \citenamefont {Sasaki},\ and\ \citenamefont {Wang}}]{Ota:2022xni}%
  \BibitemOpen
  \bibfield  {author} {\bibinfo {author} {\bibfnamefont {A.}~\bibnamefont
  {Ota}}, \bibinfo {author} {\bibfnamefont {M.}~\bibnamefont {Sasaki}}, \ and\
  \bibinfo {author} {\bibfnamefont {Y.}~\bibnamefont {Wang}},\ }\href@noop {}
  {\  (\bibinfo {year} {2022})},\ \Eprint {http://arxiv.org/abs/2211.12766}
  {arXiv:2211.12766 [astro-ph.CO]} \BibitemShut {NoStop}%
\bibitem [{\citenamefont {Gong}\ and\ \citenamefont
  {Mylova}(2022)}]{Gong:2022tfu}%
  \BibitemOpen
  \bibfield  {author} {\bibinfo {author} {\bibfnamefont {J.-O.}\ \bibnamefont
  {Gong}}\ and\ \bibinfo {author} {\bibfnamefont {M.}~\bibnamefont {Mylova}},\
  }\href {\doibase 10.1088/1475-7516/2022/07/021} {\bibfield  {journal}
  {\bibinfo  {journal} {JCAP}\ }\textbf {\bibinfo {volume} {07}},\ \bibinfo
  {pages} {021} (\bibinfo {year} {2022})},\ \Eprint
  {http://arxiv.org/abs/2202.13882} {arXiv:2202.13882 [hep-th]} \BibitemShut
  {NoStop}%
\bibitem [{\citenamefont {Born}\ and\ \citenamefont
  {Infeld}(1934)}]{Born:1934gh}%
  \BibitemOpen
  \bibfield  {author} {\bibinfo {author} {\bibfnamefont {M.}~\bibnamefont
  {Born}}\ and\ \bibinfo {author} {\bibfnamefont {L.}~\bibnamefont {Infeld}},\
  }\href {\doibase 10.1098/rspa.1934.0059} {\bibfield  {journal} {\bibinfo
  {journal} {Proc. Roy. Soc. Lond. A}\ }\textbf {\bibinfo {volume} {144}},\
  \bibinfo {pages} {425} (\bibinfo {year} {1934})}\BibitemShut {NoStop}%
\bibitem [{\citenamefont {Silverstein}\ and\ \citenamefont
  {Tong}(2004)}]{Silverstein:2003hf}%
  \BibitemOpen
  \bibfield  {author} {\bibinfo {author} {\bibfnamefont {E.}~\bibnamefont
  {Silverstein}}\ and\ \bibinfo {author} {\bibfnamefont {D.}~\bibnamefont
  {Tong}},\ }\href {\doibase 10.1103/PhysRevD.70.103505} {\bibfield  {journal}
  {\bibinfo  {journal} {Phys. Rev. D}\ }\textbf {\bibinfo {volume} {70}},\
  \bibinfo {pages} {103505} (\bibinfo {year} {2004})},\ \Eprint
  {http://arxiv.org/abs/hep-th/0310221} {arXiv:hep-th/0310221} \BibitemShut
  {NoStop}%
\bibitem [{\citenamefont {Alishahiha}\ \emph {et~al.}(2004)\citenamefont
  {Alishahiha}, \citenamefont {Silverstein},\ and\ \citenamefont
  {Tong}}]{Alishahiha:2004eh}%
  \BibitemOpen
  \bibfield  {author} {\bibinfo {author} {\bibfnamefont {M.}~\bibnamefont
  {Alishahiha}}, \bibinfo {author} {\bibfnamefont {E.}~\bibnamefont
  {Silverstein}}, \ and\ \bibinfo {author} {\bibfnamefont {D.}~\bibnamefont
  {Tong}},\ }\href {\doibase 10.1103/PhysRevD.70.123505} {\bibfield  {journal}
  {\bibinfo  {journal} {Phys. Rev. D}\ }\textbf {\bibinfo {volume} {70}},\
  \bibinfo {pages} {123505} (\bibinfo {year} {2004})},\ \Eprint
  {http://arxiv.org/abs/hep-th/0404084} {arXiv:hep-th/0404084} \BibitemShut
  {NoStop}%
\bibitem [{\citenamefont {Cai}\ and\ \citenamefont {Wang}(2010)}]{Cai:2010rt}%
  \BibitemOpen
  \bibfield  {author} {\bibinfo {author} {\bibfnamefont {Y.-F.}\ \bibnamefont
  {Cai}}\ and\ \bibinfo {author} {\bibfnamefont {Y.}~\bibnamefont {Wang}},\
  }\href {\doibase 10.1103/PhysRevD.82.123501} {\bibfield  {journal} {\bibinfo
  {journal} {Phys. Rev. D}\ }\textbf {\bibinfo {volume} {82}},\ \bibinfo
  {pages} {123501} (\bibinfo {year} {2010})},\ \Eprint
  {http://arxiv.org/abs/1005.0127} {arXiv:1005.0127 [hep-th]} \BibitemShut
  {NoStop}%
\bibitem [{\citenamefont {Cai}\ \emph {et~al.}(2011)\citenamefont {Cai},
  \citenamefont {Dent},\ and\ \citenamefont {Easson}}]{Cai:2010wt}%
  \BibitemOpen
  \bibfield  {author} {\bibinfo {author} {\bibfnamefont {Y.-F.}\ \bibnamefont
  {Cai}}, \bibinfo {author} {\bibfnamefont {J.~B.}\ \bibnamefont {Dent}}, \
  and\ \bibinfo {author} {\bibfnamefont {D.~A.}\ \bibnamefont {Easson}},\
  }\href {\doibase 10.1103/PhysRevD.83.101301} {\bibfield  {journal} {\bibinfo
  {journal} {Phys. Rev. D}\ }\textbf {\bibinfo {volume} {83}},\ \bibinfo
  {pages} {101301} (\bibinfo {year} {2011})},\ \Eprint
  {http://arxiv.org/abs/1011.4074} {arXiv:1011.4074 [hep-th]} \BibitemShut
  {NoStop}%
\bibitem [{\citenamefont {Cai}\ and\ \citenamefont {Xue}(2009)}]{Cai:2008if}%
  \BibitemOpen
  \bibfield  {author} {\bibinfo {author} {\bibfnamefont {Y.-F.}\ \bibnamefont
  {Cai}}\ and\ \bibinfo {author} {\bibfnamefont {W.}~\bibnamefont {Xue}},\
  }\href {\doibase 10.1016/j.physletb.2009.09.043} {\bibfield  {journal}
  {\bibinfo  {journal} {Phys. Lett. B}\ }\textbf {\bibinfo {volume} {680}},\
  \bibinfo {pages} {395} (\bibinfo {year} {2009})},\ \Eprint
  {http://arxiv.org/abs/0809.4134} {arXiv:0809.4134 [hep-th]} \BibitemShut
  {NoStop}%
\bibitem [{\citenamefont {Burgess}(2020)}]{Burgess:2020tbq}%
  \BibitemOpen
  \bibfield  {author} {\bibinfo {author} {\bibfnamefont {C.~P.}\ \bibnamefont
  {Burgess}},\ }\href {\doibase 10.1017/9781139048040} {\emph {\bibinfo {title}
  {{Introduction to Effective Field Theory}}}}\ (\bibinfo  {publisher}
  {Cambridge University Press},\ \bibinfo {year} {2020})\BibitemShut {NoStop}%
\bibitem [{\citenamefont {Cheung}\ \emph {et~al.}(2008)\citenamefont {Cheung},
  \citenamefont {Creminelli}, \citenamefont {Fitzpatrick}, \citenamefont
  {Kaplan},\ and\ \citenamefont {Senatore}}]{Cheung:2007st}%
  \BibitemOpen
  \bibfield  {author} {\bibinfo {author} {\bibfnamefont {C.}~\bibnamefont
  {Cheung}}, \bibinfo {author} {\bibfnamefont {P.}~\bibnamefont {Creminelli}},
  \bibinfo {author} {\bibfnamefont {A.~L.}\ \bibnamefont {Fitzpatrick}},
  \bibinfo {author} {\bibfnamefont {J.}~\bibnamefont {Kaplan}}, \ and\ \bibinfo
  {author} {\bibfnamefont {L.}~\bibnamefont {Senatore}},\ }\href {\doibase
  10.1088/1126-6708/2008/03/014} {\bibfield  {journal} {\bibinfo  {journal}
  {JHEP}\ }\textbf {\bibinfo {volume} {03}},\ \bibinfo {pages} {014} (\bibinfo
  {year} {2008})},\ \Eprint {http://arxiv.org/abs/0709.0293} {arXiv:0709.0293
  [hep-th]} \BibitemShut {NoStop}%
\bibitem [{\citenamefont {Mukhanov}\ \emph {et~al.}(1997)\citenamefont
  {Mukhanov}, \citenamefont {Abramo},\ and\ \citenamefont
  {Brandenberger}}]{Mukhanov:1996ak}%
  \BibitemOpen
  \bibfield  {author} {\bibinfo {author} {\bibfnamefont {V.~F.}\ \bibnamefont
  {Mukhanov}}, \bibinfo {author} {\bibfnamefont {L.~R.~W.}\ \bibnamefont
  {Abramo}}, \ and\ \bibinfo {author} {\bibfnamefont {R.~H.}\ \bibnamefont
  {Brandenberger}},\ }\href {\doibase 10.1103/PhysRevLett.78.1624} {\bibfield
  {journal} {\bibinfo  {journal} {Phys. Rev. Lett.}\ }\textbf {\bibinfo
  {volume} {78}},\ \bibinfo {pages} {1624} (\bibinfo {year} {1997})},\ \Eprint
  {http://arxiv.org/abs/gr-qc/9609026} {arXiv:gr-qc/9609026} \BibitemShut
  {NoStop}%
\bibitem [{\citenamefont {Green}\ and\ \citenamefont
  {Kavanagh}(2021)}]{Green:2020jor}%
  \BibitemOpen
  \bibfield  {author} {\bibinfo {author} {\bibfnamefont {A.~M.}\ \bibnamefont
  {Green}}\ and\ \bibinfo {author} {\bibfnamefont {B.~J.}\ \bibnamefont
  {Kavanagh}},\ }\href {\doibase 10.1088/1361-6471/abc534} {\bibfield
  {journal} {\bibinfo  {journal} {J. Phys. G}\ }\textbf {\bibinfo {volume}
  {48}},\ \bibinfo {pages} {043001} (\bibinfo {year} {2021})},\ \Eprint
  {http://arxiv.org/abs/2007.10722} {arXiv:2007.10722 [astro-ph.CO]}
  \BibitemShut {NoStop}%
\bibitem [{\citenamefont {Brandenberger}(2002)}]{Brandenberger:2002sk}%
  \BibitemOpen
  \bibfield  {author} {\bibinfo {author} {\bibfnamefont {R.~H.}\ \bibnamefont
  {Brandenberger}},\ }in\ \href@noop {} {\emph {\bibinfo {booktitle} {{18th IAP
  Colloquium on the Nature of Dark Energy: Observational and Theoretical
  Results on the Accelerating Universe}}}}\ (\bibinfo {year} {2002})\ \Eprint
  {http://arxiv.org/abs/hep-th/0210165} {arXiv:hep-th/0210165} \BibitemShut
  {NoStop}%
\bibitem [{\citenamefont {Sasaki}(1986)}]{Sasaki:1986hm}%
  \BibitemOpen
  \bibfield  {author} {\bibinfo {author} {\bibfnamefont {M.}~\bibnamefont
  {Sasaki}},\ }\href {\doibase 10.1143/PTP.76.1036} {\bibfield  {journal}
  {\bibinfo  {journal} {Prog. Theor. Phys.}\ }\textbf {\bibinfo {volume}
  {76}},\ \bibinfo {pages} {1036} (\bibinfo {year} {1986})}\BibitemShut
  {NoStop}%
\bibitem [{\citenamefont {Mukhanov}(1988)}]{Mukhanov:1988jd}%
  \BibitemOpen
  \bibfield  {author} {\bibinfo {author} {\bibfnamefont {V.~F.}\ \bibnamefont
  {Mukhanov}},\ }\href@noop {} {\bibfield  {journal} {\bibinfo  {journal} {Sov.
  Phys. JETP}\ }\textbf {\bibinfo {volume} {67}},\ \bibinfo {pages} {1297}
  (\bibinfo {year} {1988})}\BibitemShut {NoStop}%
\bibitem [{\citenamefont {Weinberg}(1979)}]{Weinberg:1978kz}%
  \BibitemOpen
  \bibfield  {author} {\bibinfo {author} {\bibfnamefont {S.}~\bibnamefont
  {Weinberg}},\ }\href {\doibase 10.1016/0378-4371(79)90223-1} {\bibfield
  {journal} {\bibinfo  {journal} {Physica A}\ }\textbf {\bibinfo {volume}
  {96}},\ \bibinfo {pages} {327} (\bibinfo {year} {1979})}\BibitemShut
  {NoStop}%
\bibitem [{\citenamefont {Polchinski}(1984)}]{Polchinski:1983gv}%
  \BibitemOpen
  \bibfield  {author} {\bibinfo {author} {\bibfnamefont {J.}~\bibnamefont
  {Polchinski}},\ }\href {\doibase 10.1016/0550-3213(84)90287-6} {\bibfield
  {journal} {\bibinfo  {journal} {Nucl. Phys. B}\ }\textbf {\bibinfo {volume}
  {231}},\ \bibinfo {pages} {269} (\bibinfo {year} {1984})}\BibitemShut
  {NoStop}%
\bibitem [{\citenamefont {Polchinski}(1992)}]{Polchinski:1992ed}%
  \BibitemOpen
  \bibfield  {author} {\bibinfo {author} {\bibfnamefont {J.}~\bibnamefont
  {Polchinski}},\ }in\ \href@noop {} {\emph {\bibinfo {booktitle} {{Theoretical
  Advanced Study Institute (TASI 92): From Black Holes and Strings to
  Particles}}}}\ (\bibinfo {year} {1992})\ pp.\ \bibinfo {pages} {0235--276},\
  \Eprint {http://arxiv.org/abs/hep-th/9210046} {arXiv:hep-th/9210046}
  \BibitemShut {NoStop}%
\bibitem [{\citenamefont {Georgi}(1993)}]{Georgi:1993mps}%
  \BibitemOpen
  \bibfield  {author} {\bibinfo {author} {\bibfnamefont {H.}~\bibnamefont
  {Georgi}},\ }\href {\doibase 10.1146/annurev.ns.43.120193.001233} {\bibfield
  {journal} {\bibinfo  {journal} {Ann. Rev. Nucl. Part. Sci.}\ }\textbf
  {\bibinfo {volume} {43}},\ \bibinfo {pages} {209} (\bibinfo {year}
  {1993})}\BibitemShut {NoStop}%
\bibitem [{\citenamefont {Weinberg}(2021)}]{Weinberg:2021exr}%
  \BibitemOpen
  \bibfield  {author} {\bibinfo {author} {\bibfnamefont {S.}~\bibnamefont
  {Weinberg}},\ }\href {\doibase 10.1140/epjh/s13129-021-00004-x} {\bibfield
  {journal} {\bibinfo  {journal} {Eur. Phys. J. H}\ }\textbf {\bibinfo {volume}
  {46}},\ \bibinfo {pages} {6} (\bibinfo {year} {2021})},\ \Eprint
  {http://arxiv.org/abs/2101.04241} {arXiv:2101.04241 [hep-th]} \BibitemShut
  {NoStop}%
\bibitem [{\citenamefont {Green}(2022)}]{Green:2022ovz}%
  \BibitemOpen
  \bibfield  {author} {\bibinfo {author} {\bibfnamefont {D.}~\bibnamefont
  {Green}},\ }\href@noop {} {\  (\bibinfo {year} {2022})},\ \Eprint
  {http://arxiv.org/abs/2210.05820} {arXiv:2210.05820 [hep-th]} \BibitemShut
  {NoStop}%
\bibitem [{\citenamefont {Weinberg}(2008)}]{Weinberg:2008hq}%
  \BibitemOpen
  \bibfield  {author} {\bibinfo {author} {\bibfnamefont {S.}~\bibnamefont
  {Weinberg}},\ }\href {\doibase 10.1103/PhysRevD.77.123541} {\bibfield
  {journal} {\bibinfo  {journal} {Phys. Rev. D}\ }\textbf {\bibinfo {volume}
  {77}},\ \bibinfo {pages} {123541} (\bibinfo {year} {2008})},\ \Eprint
  {http://arxiv.org/abs/0804.4291} {arXiv:0804.4291 [hep-th]} \BibitemShut
  {NoStop}%
\bibitem [{\citenamefont {Gleyzes}\ \emph {et~al.}(2013)\citenamefont
  {Gleyzes}, \citenamefont {Langlois}, \citenamefont {Piazza},\ and\
  \citenamefont {Vernizzi}}]{Gleyzes:2013ooa}%
  \BibitemOpen
  \bibfield  {author} {\bibinfo {author} {\bibfnamefont {J.}~\bibnamefont
  {Gleyzes}}, \bibinfo {author} {\bibfnamefont {D.}~\bibnamefont {Langlois}},
  \bibinfo {author} {\bibfnamefont {F.}~\bibnamefont {Piazza}}, \ and\ \bibinfo
  {author} {\bibfnamefont {F.}~\bibnamefont {Vernizzi}},\ }\href {\doibase
  10.1088/1475-7516/2013/08/025} {\bibfield  {journal} {\bibinfo  {journal}
  {JCAP}\ }\textbf {\bibinfo {volume} {08}},\ \bibinfo {pages} {025} (\bibinfo
  {year} {2013})},\ \Eprint {http://arxiv.org/abs/1304.4840} {arXiv:1304.4840
  [hep-th]} \BibitemShut {NoStop}%
\bibitem [{\citenamefont {Cabass}\ \emph {et~al.}(2023)\citenamefont {Cabass},
  \citenamefont {Ivanov}, \citenamefont {Lewandowski}, \citenamefont
  {Mirbabayi},\ and\ \citenamefont {Simonovi\'c}}]{Cabass:2022avo}%
  \BibitemOpen
  \bibfield  {author} {\bibinfo {author} {\bibfnamefont {G.}~\bibnamefont
  {Cabass}}, \bibinfo {author} {\bibfnamefont {M.~M.}\ \bibnamefont {Ivanov}},
  \bibinfo {author} {\bibfnamefont {M.}~\bibnamefont {Lewandowski}}, \bibinfo
  {author} {\bibfnamefont {M.}~\bibnamefont {Mirbabayi}}, \ and\ \bibinfo
  {author} {\bibfnamefont {M.}~\bibnamefont {Simonovi\'c}},\ }\href {\doibase
  10.1016/j.dark.2023.101193} {\bibfield  {journal} {\bibinfo  {journal} {Phys.
  Dark Univ.}\ }\textbf {\bibinfo {volume} {40}},\ \bibinfo {pages} {101193}
  (\bibinfo {year} {2023})},\ \Eprint {http://arxiv.org/abs/2203.08232}
  {arXiv:2203.08232 [astro-ph.CO]} \BibitemShut {NoStop}%
\bibitem [{\citenamefont {Peskin}\ and\ \citenamefont
  {Schroeder}(1995)}]{Peskin:1995ev}%
  \BibitemOpen
  \bibfield  {author} {\bibinfo {author} {\bibfnamefont {M.~E.}\ \bibnamefont
  {Peskin}}\ and\ \bibinfo {author} {\bibfnamefont {D.~V.}\ \bibnamefont
  {Schroeder}},\ }\href@noop {} {\emph {\bibinfo {title} {{An Introduction to
  quantum field theory}}}}\ (\bibinfo  {publisher} {Addison-Wesley},\ \bibinfo
  {address} {Reading, USA},\ \bibinfo {year} {1995})\BibitemShut {NoStop}%
\end{thebibliography}%
\end{document}